\documentclass{aa}  

\usepackage{graphicx}

\usepackage{txfonts}
\usepackage{natbib}
\bibpunct{(}{)}{;}{a}{}{,} 

\begin{document} 

   \title{Understanding post-red giant branch binaries through stable mass transfer}

   \author{C. A. S. Moltzer\inst{1},
          O. R. Pols\inst{1},
          H. Van Winckel\inst{2},
          K. D. Temmink\inst{1},
          \and
          M. W. Wijdeveld\inst{1} 
          }
    
   \institute{Department of Astrophysics/IMAPP, Radboud University Nijmegen, PO Box 9010, 6500 GL Nijmegen, The Netherlands\\
              \email{casper.moltzer@ru.nl}
         \and
             Institute of Astronomy, KU Leuven, Celestijnenlaan 200D, 3001 Leuven, Belgium\\
             }

   \date{Received 16 July 2025 / Accepted 3 September 2025}

  \abstract
   {Post-red giant branch (post-RGB) and post-asymptotic giant branch (post-AGB) binaries consist of a primary star that has recently evolved off either the RGB or AGB after losing the majority of its envelope and a main-sequence companion. They are distinguished by having luminosities below and above the tip of the RGB, respectively. These systems are characterised by the presence of a stable, dusty circumbinary disc, identified by a near-IR excess. Observed Galactic post-AGB and post-RGB binaries have orbital periods and eccentricities that are at odds with binary population synthesis models.}
   {In this work, we focus on post-RGB binaries. We investigate whether stable mass transfer can explain the orbital periods of such binaries by comparing stable mass transfer models with the known sample of 38 Galactic post-RGB binaries.}
   {We systematically determined the luminosities of the Galactic post-RGB and post-AGB binary sample using spectral energy distribution fitting. We computed evolution models for low- and intermediate-mass binaries with RGB donors at two metallicities using the detailed stellar evolution code, MESA. We selected the stable mass transfer models that result in primaries with effective temperatures within the observed range of post-RGB binaries ($4000-8500$ K).}
   {From our model grids, we find that low-mass post-RGB binaries are expected to follow strict luminosity-orbital period relations. The Galactic post-RGB binaries appear consistent with these luminosity-orbital period relations if we assume that their orbits remained eccentric during mass transfer and that the donor star filled its Roche lobe at periastron. However, our models are unable to explain the eccentricities themselves. Furthermore, the post-mass-transfer ages of observed post-RGB binaries estimated using our models are significantly longer than the predicted dissipation timescales of their circumbinary discs.}
   {The stable mass transfer formation channel appears to explain the orbital periods of Galactic post-RGB binaries. This formation scenario could be tested more extensively by obtaining the orbits of additional Galactic systems, as well as those of the numerous candidates in the Magellanic Clouds, through long-term radial velocity monitoring. Additionally, we expect that Gaia Data Release 4 will improve the luminosities of Galactic post-RGB binaries, which will allow for a more accurate comparison with post-RGB luminosity-orbital period relations.}


    \authorrunning{C. A. S. Moltzer et al.}
    \titlerunning{Understanding post-RGB binaries through stable mass transfer}
   \maketitle

\section{Introduction}
Post-asymptotic giant branch (post-AGB) stars are low- to intermediate-mass stars that are in a transitional phase between the AGB and the white dwarf stages \citep[for comprehensive reviews, see][]{VanWinckel2003,VanWinckel2019}. These evolved stars have recently lost their envelopes and are contracting towards higher effective temperatures while maintaining a constant luminosity \citep[e.g.][]{MillerBertolami2016}. 

Optically bright post-AGB stars are observed to have an IR excess in their spectral energy distributions (SEDs) due to the presence of circumstellar dust. About one third of the Galactic post-AGB stars exhibit a specific type of SED with a near-IR excess component, which is interpreted as a warm dusty disc around the object, likely formed during binary interaction \citep[e.g.][]{VanWinckel2019,VanWinckel2025}. Indeed, these post-AGB stars with disc-type SEDs are observed in binaries with orbital periods ranging from 100 to 3000 days, and with eccentricities from 0 up to $0.63$ \citep{Oomen2018, Kluska2022}. 

Post-AGB binaries are ideal candidates for testing binary evolution. The short lifetime of the post-AGB phase, ranging from $10^3$ to $10^5$ years \citep{MillerBertolami2016}, indicates that the primary star has recently lost its envelope presumably through mass transfer. The circumbinary discs of these systems are also indicative of recent binary interaction, as the estimated dissipation timescales of these discs are $5\times10^3-4\times10^4$ years \citep[e.g.][]{Bujarrabal2017,Izzard2023}.

Furthermore, the envelope loss experienced by the primary star is expected to be the first binary interaction in post-AGB binaries. The companions in these systems, although not detected in the SEDs, are likely to be main-sequence stars. This is based on their broad inferred mass distribution, which lacks the narrow peak at about 0.6 $M_{\odot}$ that would be expected if the companions were predominantly white dwarfs \citep{Oomen2018}. Additionally, modelling of the collimated outflows observed in some of these systems has shown that the jet velocities are comparable to the escape velocity of a main-sequence star, and about one order of magnitude lower than the escape velocity of a white dwarf \citep{Bollen2022}.

The orbital properties of the Galactic post-AGB binaries appear inconsistent with binary population synthesis models \citep[see discussion in][]{Oomen2018}. Their orbital periods are too short to have contained a giant star without undergoing mass transfer via Roche-lobe overflow (RLOF). However, the canonical view of mass transfer involving a giant donor star is that it is unstable because the radius of the donor is expected to increase in response to mass loss on account of its convective envelope \citep[e.g.][]{Webbink1985}. This would lead to a common envelope, which causes the in-spiral of the binary, shortening the orbital period to less than 100 days and possibly resulting in a merger \citep[e.g.][]{Nie2012}. 

The eccentric orbits of many post-AGB binaries are also difficult to reconcile with binaries that have undergone RLOF, since tidal interactions are expected to have circularised the orbits \citep{Hurley2002, Oomen2020}. This long-standing eccentricity-orbital period problem is not unique to post-AGB binaries, as it is also observed in other post-mass-transfer low- to intermediate-mass binary systems such as wide subdwarf B binaries \citep[e.g.][]{Vos2017,Vos2019}, carbon-enhanced metal-poor stars with s-process enrichment \citep[e.g.][]{Jorissen2016, Hansen2016}, barium and CH stars \citep[e.g.][]{Jorissen2019, Escorza2019}, extrinsic S-type stars \citep[e.g.][]{Fekel2000, Jorissen2019}, and long-period binary central stars of planetary nebulae \citep[e.g.][]{VanWinckel2014,Jones2017}.

Some objects classified as post-AGB stars have luminosities below the tip of the red giant branch (RGB; about 2500 $L_{\odot}$) and are interpreted as post-RGB stars \citep{Kamath2014,Kamath2015,Kamath2016}. Post-RGB stars can only form through binary interactions because, unlike AGB stars, RGB stars lack the mass-loss rates required to shed the majority of their envelopes. Post-RGB stars are the precursor phase of helium white dwarfs, whereas post-AGB stars result in carbon-oxygen white dwarfs. Related to post-RGB stars are subdwarf B binaries, which are thought to form through mass transfer when the donor star was near the tip of the RGB \citep[e.g.][]{Han2002,Han2003,Vos2019}. Luminosities of post-RGB and post-AGB stars can be determined from SED fitting combined with a distance measurement \citep{Oomen2018, Kluska2022}, although they have large uncertainties due to substantial errors on both the distance measurement and the reddening determination.

Recently, \citet{Temmink2023} computed detailed binary evolution models, which show that mass transfer involving RGB donors can be stable for a substantial portion of the binary parameter space, confirming earlier results \citep{Han2002,Ge2010,Ge2020,Ge2020b}. Such binaries are expected to follow a strict relation between the maximum radius reached by the donor star during mass transfer and the orbital period at the end of mass transfer. This is because during stable mass transfer the radius of the donor star is approximately equal to its Roche lobe radius. Moreover, these RGB stars are expected to exhibit tight relations between their core mass, radius, and luminosity. As a consequence, post-RGB binaries are expected to follow a luminosity-orbital period relation at the end of a stable mass transfer episode \citep[e.g.][]{Webbink1983}.

In this paper we investigate whether the properties of the Galactic post-RGB binary sample can be explained by stable mass transfer models with RGB donors. In particular, we determine the luminosity-orbital period relation from the post-RGB models computed by \citet{Temmink2023} and compare them with the orbital periods and luminosities of Galactic post-RGB binaries.

In Sect. \ref{data} we present the sample of Galactic post-RGB and post-AGB binaries, along with the samples of post-RGB and post-AGB binary candidates in the Large Magellanic Cloud (LMC) and Small Magellanic Cloud (SMC). In Sect. \ref{models} we present our full grid of models that have undergone stable mass transfer. In Sect. \ref{model results} we classify these models into several object types and study their properties, in particular the luminosity-orbital period relations for post-RGB binaries. In Sect. \ref{comparison to data} we compare the orbital periods and luminosities of the sample of Galactic post-RGB binaries with these luminosity-orbital period relations. We discuss our results in Sect. \ref{discussion} and present our conclusions in Sect. \ref{conclusion}.

\section{Observational data} \label{data}

\subsection{Galactic sample} \label{Galactic data}
We defined the Galactic sample as the 85 post-AGB and post-RGB binaries from \citet{Kluska2022}. We cross-referenced these objects with other studies to supplement any missing orbital periods or eccentricities. These orbital properties were determined either spectroscopically \citep[e.g.][]{Oomen2018, Manick2019}, which yields both the orbital period and the eccentricity, or photometrically \citep[e.g.][]{Kiss2007, Percy2015}, which yields only the orbital period. In the latter case, in addition to any stellar pulsations of the post-AGB or post-RGB star that are usually observed \citep{Wallerstein2002}, a long-term modulation was detected that is interpreted as the periodic shrouding of the star by the circumbinary disc \citep[e.g.][and references therein]{Kiss2017}.

\begin{figure}
    \resizebox{\hsize}{!}{\includegraphics{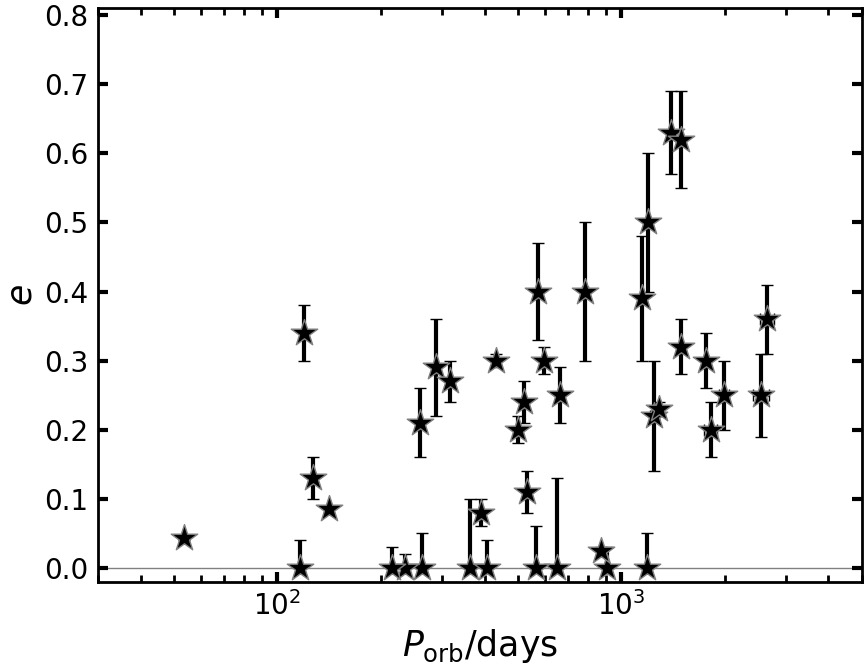}}
    \caption{Eccentricity-orbital period diagram of the Galactic post-AGB and post-RGB binary sample.}
    \label{figure:eP}
\end{figure}

This resulted in a total of 53 objects with known orbits. Of these, 39 have spectroscopically determined orbital periods and eccentricities. The remaining 14 have only photometrically determined long secondary periods, which we assume to be orbital periods. The systems, along with their orbital properties and corresponding references, are presented in Table \ref{table:orbits}. The orbital periods of the Galactic sample range from 53 days to 2654 days, and their eccentricities range between 0 and 0.63. Fig. \ref{figure:eP} shows the eccentricity-orbital period diagram of Galactic post-AGB and post-RGB binaries, which has been updated from \citet{Oomen2018} to include the additional systems from this sample. 

We determined the bolometric fluxes and effective temperatures of these objects from SED fitting using the methods described in \citet{Oomen2018} and \citet{Kluska2022}. To convert these fluxes to luminosities, we used the distances from \citet{BailerJones2021}, who in turn used data from Gaia Early Data Release 3 \citep[hereafter EDR3]{Gaia2021}. Specifically, we used the geometric distances instead of the typically more precise photogeometric distances. The former uses solely the parallax data from Gaia EDR3, together with a distance prior, while the latter also uses the colour and apparent magnitude data, together with a colour-magnitude prior. Since these objects are unresolved binaries within the Gaia EDR3 data, and the colour-magnitude prior from \citet{BailerJones2021} is only valid for single stars, it is more appropriate to use the geometric distances. 

In some cases the IR luminosity, $L_\mathrm{IR}$ (the luminosity obtained by integrating over the IR excess), exceeds the stellar luminosity, $L_*$ (the photospheric luminosity obtained by integrating over the model photosphere fitted to the dereddened optical fluxes). In these cases the dereddening in the line-of-sight is insufficient to explain the IR excess, which indicates an edge-on disc. This causes the optical flux to be dominated by scattered light rather than direct reddened light, such that the fitted stellar luminosity is significantly lower than the true stellar luminosity. To correct for this, the IR luminosity was used as a proxy for the stellar luminosity for objects with an IR-to-stellar luminosity ratio $L_\mathrm{IR}/L_*$ greater than $1.0$. It should be noted that the luminosities can have significant uncertainties due to both the uncertainty in the distances from \citet{BailerJones2021} and the uncertainty in the reddening determinations $E(B-V)$ \citep{Kluska2022,Menon2024}. The SED fitting data for the sample are presented in Table \ref{table:GalacticSEDs}.

\begin{figure}
    \resizebox{\hsize}{!}{\includegraphics{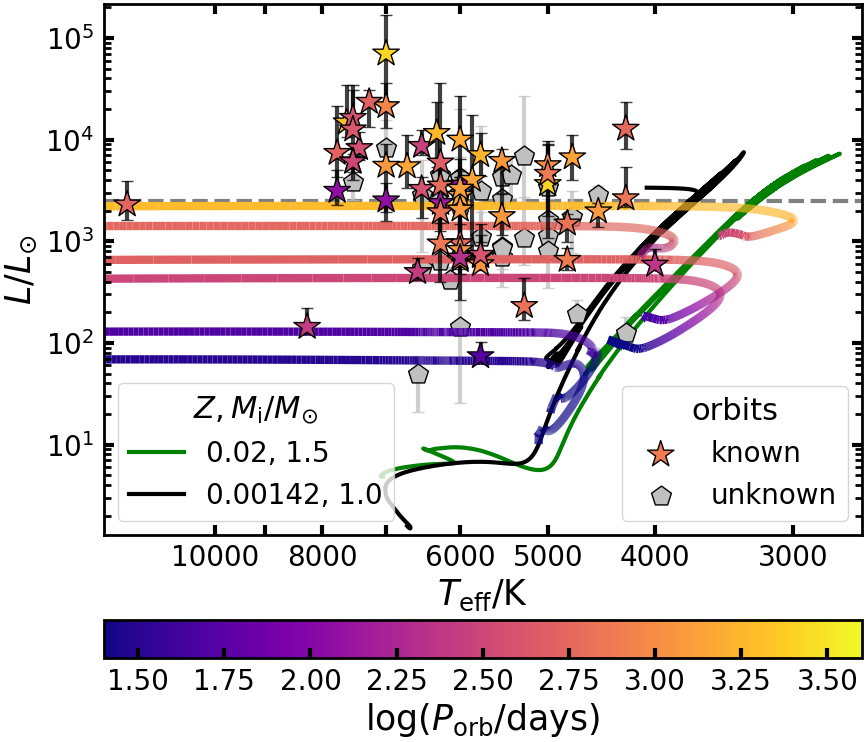}}
    \caption{HRD of the Galactic post-AGB and post-RGB binary sample. Different markers denote whether the orbital periods of the objects are known. The $T_\mathrm{eff}$ error bars of $\pm250$ K were omitted for the sake of clarity. Evolutionary tracks of our single-star models with $Z=0.02$, $M_\mathrm{i}=1.5$ $M_{\odot}$ and $Z=0.00142$, $M_\mathrm{i}=1.0$ $M_{\odot}$ are shown in green and black, respectively, up to the final AGB evolution. In addition, we show a few illustrative post-RGB evolution tracks with initial orbital periods ranging from 3.95 days to 258 days (from bottom to top). These models have donors identical to their starting point on the corresponding single-star evolutionary track and have mass ratios equal to unity prior to mass transfer. Orbital periods of the sample binaries as well as the orbital period evolution of the models are shown by the colour bar. The dashed grey line corresponds to the RGB-tip luminosity of 2500 $L_{\odot}$.}
    \label{figure:HRD_Galactic}
\end{figure}

Fig. \ref{figure:HRD_Galactic} shows the sample stars in the Hertzsprung-Russell diagram (HRD), colour-coded according to their orbital periods. The effective temperatures of the objects range between 4000 K and 8500 K, along with one object (HD 137569) having an effective temperature of 12000 K. The luminosities range from 70 $L_{\odot}$ to 70000 $L_{\odot}$, with 38 of the 85 objects having luminosities below the RGB-tip luminosity, which we define\footnote{This value is taken from \citet{Kamath2015}, who define it in relation to the Magellanic Clouds. We use this as an approximation for the Galactic objects, because of their spread in metallicity.} as approximately 2500 $L_{\odot}$.

The metallicity of a star is an important parameter that influences its evolution. However, for post-AGB and post-RGB binaries, iron-to-hydrogen abundance ratio ([Fe/H]) measurements are in most cases not indicative of the metallicity ($Z$) of the object due to depletion. This phenomenon is caused by accretion of gas from the circumbinary disc, which has become devoid of refractory elements by dust formation in the circumstellar environment \citep{Waters1992,Oomen2019}. For these depleted objects, the zinc-to-hydrogen abundance ratio [Zn/H] is used instead as a metallicity tracer of the main-sequence composition of the star ([M/H]), because the lower condensation temperature of Zn in oxygen-rich circumstellar environments makes it much less susceptible to depletion \citep[e.g.][and references therein]{Giridhar2005,Mohorian2025}. In the absence of [Zn/H] data for a given depleted object, the sulphur-to-hydrogen abundance ratio [S/H] is used as [M/H], as it has a comparably low condensation temperature. In the sample, [M/H] ranges from $\mathrm{[Zn/H]}=-1.3$ (HR~4049) to $\mathrm{[S/H]}=0.4$ (AR~Pup and HD~213985). The mean [M/H] of the sample is $-0.4$, with a standard deviation of 0.4. This excludes objects for which no depletion information is available, as well as objects lacking both [Zn/H] and [S/H] despite exhibiting depletion. These metallicity values appear consistent with the assumption that these objects are associated with the Galactic thick disc \citep[e.g.][and references therein]{Kluska2022}, which has a mean [Fe/H] of approximately $-0.30$ \citep{Sharma2019}.

\subsection{LMC and SMC samples} \label{MC data}
We constructed the LMC sample of 101 post-AGB and post-RGB binary candidates by taking 38 objects with disc-type SEDs\footnote{The SED types are less well constrained in the LMC and SMC because the Spitzer IR survey had good sensitivity only up to 24 $\mu$m.} from \citet{VanAarle2011}, and supplementing them with 55 objects from \citet{Kamath2015} and eight objects from \citet{Manick2018} with disc-type SEDs. Similarly, we constructed the SMC sample of post-AGB and post-RGB binary candidates by taking all 27 objects with disc-type SEDs from \citet{Kamath2014}. Using the same SED fitting method as for the Galactic objects, we determined their luminosities and effective temperatures, which are presented in Table \ref{table:MCsSEDs}. The distances used were 50 kpc for the LMC \citep{Walker2012,Pietrzynski2013} and 62 kpc for the SMC \citep{Graczyk2013}. As these distances are well determined, the uncertainties on the luminosities of the LMC and SMC objects are smaller and dominated by the uncertainties in the reddening determination. 

Two objects in the LMC sample, OGLE-LMC-T2CEP-032 and OGLE-LMC-T2CEP-200, have photometric orbital periods of 916 days and 850 days, respectively \citep{Manick2018}. Additionally, one object in the SMC sample, OGLE-SMC-T2CEP-018, has a photometric orbital period of 1404 days \citep{Groenewegen2017}.

\begin{figure}
    \resizebox{\hsize}{!}{\includegraphics{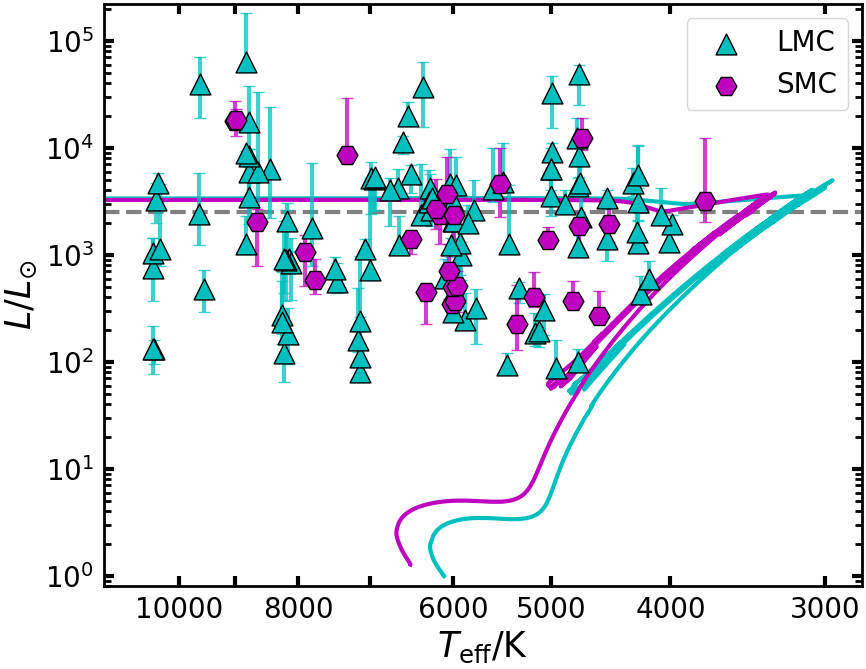}}
    \caption{HRD of the post-AGB and post-RGB binary candidate samples in the LMC and SMC. The $T_{\mathrm{eff}}$ error bars of ±250 K were omitted for the sake of clarity. Also shown are evolutionary tracks of single star models, taken from the MIST database \citep{Choi2016}, with $M_i/M_{\odot}=1.0$ and the [Fe/H] of $-0.30$ and $-0.65$ corresponding to the mean metallicities of the LMC and SMC, respectively \citep{Westerlund1997,Larsen2000}. The dashed grey line corresponds to the RGB-tip luminosity of 2500 $L_{\odot}$.}
    \label{figure:HRD_MCs}
\end{figure}

Fig. \ref{figure:HRD_MCs} shows the LMC and SMC samples plotted in the HRD. The effective temperatures of the objects span a range of $3800-11000$ K. The luminosities of the LMC sample range between 80 $L_{\odot}$ and 64000 $L_{\odot}$, with 56 out of 101 objects having luminosities below the RGB-tip luminosity. The luminosities of the SMC sample range between 220 $L_{\odot}$ and 18000 $L_{\odot}$, with 19 out of 27 objects having luminosities below the RGB-tip luminosity.

\section{Stable mass transfer models} \label{models}

\subsection{Binary model grids}
For the bulk of our work, we used the detailed binary models from \citet{Temmink2023}, which assume conservative mass transfer and circular orbits. This grid of models covers the stable mass transfer regime and contains primaries that evolve through post-RGB or post-AGB phases. The metallicity is slightly super-solar ($Z=0.02$), which corresponds to [Fe/H] of $0.15$ using $\mathrm{[Fe/H]} = \log(Z/Z_{\odot})$, where $Z_{\odot}$ is the solar metallicity taken to be $0.0142$ \citep{Asplund2009}. 

The models of \citet{Temmink2023} encompass low- and intermediate-mass primaries with initial masses ($M_\mathrm{i}$) of $1.0$, $1.5$, $2.0$, $2.5$, $3.0$, $4.0$, $5.0$, $6.0$, $7.0$, $8.0$ $M_{\odot}$. For each of these primaries, RLOF was initiated for a grid of orbital separations (spaced logarithmically by approximately 0.2 dex) and mass ratios\footnote{\citet{Temmink2023} provides models with mass ratios greater than 1.0. However, we did not use these because the companions of post-RGB binaries are most likely main-sequence stars that are initially less massive than the primaries \citep{Oomen2018}.} ($q=M_\mathrm{a}/M_\mathrm{d}$, ranging from 0.1 to 1.0 in steps of 0.1), such that the primary filled its Roche lobe at a number of starting points along the single-star evolution tracks of these primaries, selected between the end of the main sequence and the first thermal pulse (TP) during the AGB. The companion stars were treated as point masses. Further details concerning these models can be found in \citet{Temmink2023}. 

In order to obtain a higher density of stable mass transfer models for this grid in the low primary mass regime that result in post-RGB binaries, we computed additional models for $M_i$ of $1.25$ $M_{\odot}$ and $1.75$ $M_{\odot}$. We used a method identical to that discussed in \citet{Temmink2023}, utilising version r12115 of MESA \citep{Paxton2011,Paxton2013,Paxton2015,Paxton2018,Paxton2019}. These binary models cover a smaller parameter space than those of \citet{Temmink2023}, as we selected donors that undergo mass transfer between about halfway up the RGB and the tip of the RGB. The additional models in combination with the models of \citet{Temmink2023} are referred to as the solar models.

Moreover, to understand the effect of metallicity on the formation of post-RGB and post-AGB binaries, we computed a grid of models that are identical\footnote{Because of an incorrect code configuration \citep{Temmink2025}, the models of \citet{Temmink2023} are not truly conservative. This has been fixed for this new model grid.} to those of \citet{Temmink2023}, except that the metallicity was set to $\log(Z/Z_{\odot})=-1.0$ and $q$ was incremented in steps of $0.2$ instead of $0.1$. These models with sub-solar metallicity are referred to as the metal-poor models.

\subsection{Model selection} \label{selection}
From the two grids, all models were selected in which mass transfer was stable according to all three of the criteria described in \citet{Temmink2023}. Firstly, the quasi-adiabatic criterion ascertains that the surface layers of the donor can thermally readjust before being stripped, thus preventing unstable adiabatic mass transfer. Furthermore, our selection ascertains that mass transfer does not occur on the dynamical timescale of the donor, and that the radius of the donor does not exceed the outer Roche lobe radius of the system. The initial parameters of our stable mass transfer models can be found in Tables \ref{table:solarmodels} and  \ref{table:mpmodels} for the solar and metal-poor grids, respectively.

Moreover, we selected the models from our grid for which the primary star resembles the observed stars in the sample after mass transfer. Firstly, we selected the post-mass-transfer models from each model sequence. This was done by identifying the point at which the mass-transfer rate first falls below $10^{-12}$ $M_{\odot}$/years, and selecting all subsequent models with a mass-transfer rate below this value. From these models, we then selected those for which the effective temperature of the primary star falls within the observed range of the sample ($4000-8500$ K). 

Additionally, as the sample binaries have undetected companions, we excluded the models for which the luminosity of the secondary star exceeds $5\%$ of the luminosity of the primary star. Since the secondary star was treated as a point mass in our models, we estimated its luminosity from its mass. We used the mass of the secondary at the onset of mass transfer to obtain a lower limit on the luminosity of the secondary star, assuming that the companion does not accrete any mass. We estimated this minimum luminosity from zero-age main-sequence (ZAMS) mass-luminosity relations for both metallicities using initially non-rotating models taken from the MESA Isochrones \& Stellar Tracks (MIST) database \citep{Choi2016}.  This selection eliminates models for which we expect the companion to be detected, regardless of uncertainties in binary evolution. In particular, the mass-transfer efficiency of the sample binaries is currently unknown, which means that the post-interaction secondary masses are potentially overestimated by our conservative mass transfer models (see Sect. \ref{discussion} for further details).

In a few cases, our selection criteria yield multiple disjunct model sequences from the same binary evolution track (for example, additional phases of mass transfer can occur due to a late TP on the AGB \citep[e.g.][]{Lawlor2023}, or when RLOF starts on the RGB but is interrupted by a horizontal branch phase before continuing on the AGB). When presenting our results, we avoided over-representing these particular binary evolution tracks by not depicting all their model sequences. Instead, we selected the model sequence that most closely matches the constant luminosity evolution to higher effective temperatures observed for post-AGB and post-RGB stars in the HRD. Firstly, the sequences that increase in effective temperature were selected. Secondly, the sequence that covers the largest range of effective temperatures was selected. If multiple sequences meet both these two additional criteria, the one with the smallest difference between its minimum and maximum luminosity was selected.

\section{Results of post-mass-transfer models} \label{model results}

\subsection{Types of post-mass-transfer donors}
\begin{figure*}
\centering
   \includegraphics[width=18cm]{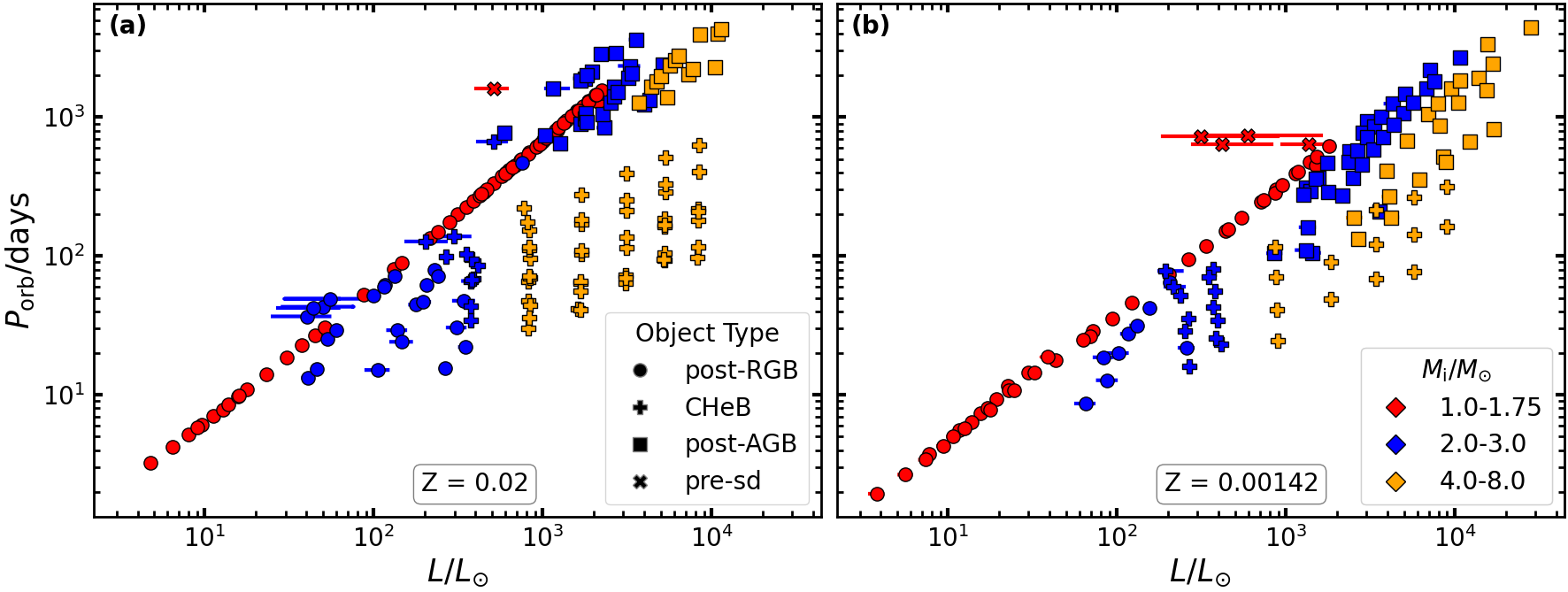}
     \caption{Orbital period-luminosity diagram of the donor stars in our models following stable mass transfer with effective temperatures in the range of $4000-8500$ K, colour-coded by initial primary mass. Panel (a) corresponds to the solar metallicity grid, and panel (b) to the metal-poor grid. For those models that do not exhibit constant luminosity after mass transfer, the lines depict the variation in luminosity as the model traverses the effective temperature range, with the markers denoting the time-weighted average of the corresponding evolution. Different markers denote the type of post-mass-transfer donor: post-RGB, central He-burning, post-AGB, or pre-subdwarf (pre-sd).}
     \label{figure:models}
\end{figure*}

Fig. \ref{figure:models} shows the luminosities and orbital periods of our selected post-mass-transfer binary models. We distinguished four distinct types of post-mass-transfer donors. Objects with a central He mass fraction of less than $10^{-4}$ were classified as post-AGB stars. For objects that still have helium in their cores, we checked whether the He-burning luminosity exceeded 10 $L_{\odot}$ during their prior evolution, which indicates that He ignition has occurred \citep{Arancibia-Roja2024}. If this was not the case, the objects were classified as post-RGB stars. In the case of objects where helium ignition occurred, we checked whether the central helium mass fraction decreased by more than $10^{-4}$ below the initial mass fraction of the helium core (i.e. $1-Z$). If this was the case, we classified these objects as stripped central helium burning (CHeB) stars. If this was not the case, the objects were classified as pre-subdwarfs. These stars were presumed to have undergone off-centre helium ignition, a phenomenon typical of subdwarf progenitors. We expect these stars to contract to higher effective temperatures, probably exceeding 8500 K, after which the steady phase of central He-burning as a hot subdwarf commences \citep{Arancibia-Roja2024}. 

The colours in Fig. \ref{figure:models} indicate the initial mass of the model's primary star. The majority of post-RGB stars have initial primary masses below 2.0 $M_{\odot}$ and exhibit a tight luminosity-orbital period relation. This results from the strict core mass-luminosity and core mass-radius relations followed by low-mass stars that develop degenerate cores on the RGB (see Sect. \ref{relation} for details). Since the initial mass function is weighted towards low-mass progenitors, we expect the post-RGB binaries that follow the luminosity-orbital period relations to dominate over any binaries originating from higher initial mass primaries.

A subset of post-RGB stars, characterised by initial primary masses ranging from 2.0 $M_{\odot}$ to 3.0 $M_{\odot}$ and luminosities below 300 $L_{\odot}$, deviate from the post-RGB luminosity-orbital period relations discussed above. This deviation can be attributed to the non-degenerate cores of these stars. Our models show that many of these stars will undergo helium ignition, but only for effective temperatures above the chosen range of $4000-8500$ K. This means that this subset of post-RGB stars are precursors of central He-burning stars. 

The few pre-subdwarf models, all with initial primary masses below 2.0 $M_{\odot}$, have orbital periods slightly longer than the maximum orbital periods reached by the post-RGB models, which are about 1600 days for $Z=0.02$ and 630 days for $Z=0.00142$. Their luminosities vary significantly as they evolve through the effective temperature range of $4000-8500$ K, resulting in an extended horizontal track in Fig. \ref{figure:models}. The narrow range of orbital periods of these models is attributed to the fact that mass transfer should have started when the primary was sufficiently close to the tip of the RGB for the degenerate core to undergo the helium flash. In these objects, helium ignition occurs as a series of flashes, each leading to a contraction to higher effective temperatures and a decrease in luminosity \citep{Arancibia-Roja2024}, thereby explaining the large variations in luminosity exhibited by our pre-subdwarf models. We find that the time taken by the pre-subdwarf models to evolve through the selected effective temperature range is considerably shorter, by a factor between 2.5 and 100, than for post-RGB stars of similar core mass. This implies that pre-subdwarfs are expected to be exceedingly rare in the effective temperature range of $4000-8500$ K compared to post-RGB stars. 

The CHeB stars in Fig. \ref{figure:models} have shorter orbital periods compared to post-RGB or post-AGB models of equivalent core mass. These models have undergone mass transfer from a Hertzsprung gap or subgiant donor, which have considerably smaller radii than giant stars, resulting in shorter orbital periods. These donors have radiative envelopes, leading to mass transfer occurring on thermal to near-nuclear timescales, which are substantially slower than for giant donors. This results in a lower mass-transfer rate, which in theory would allow for a higher mass-transfer efficiency because the companion star has more time to accrete the transferred material, allowing it to grow considerably in mass. We expect these more massive companions to have luminosities closer to those of the CHeB star. The absence of a detectable companion for all objects in the sample therefore makes it improbable that these CHeB objects will be identified as post-RGB or post-AGB binaries. Additionally, it remains unclear whether binaries with Hertzsprung gap or subgiant donors would harbour dusty discs similar to those observed for the sample objects. Dust formation is sensitive to temperature \citep{Hofner2018}, and the higher effective temperature of these donors compared to those of giant donors could inhibit this process. 

Among the post-AGB binaries in Fig. \ref{figure:models}, there is significantly more scatter in their orbital periods compared to the post-RGB binaries. Note that our model grids do not include mass transfer during the TPAGB, so the model predictions for post-AGB binaries are incomplete. We find that post-AGB binaries overlap with post-RGB binaries in the luminosity regime 600 $L_{\odot}$ to 2300 $L_{\odot}$. Moreover, Fig. \ref{figure:models} shows that the post-AGB binaries in the metal-poor grid have orbital periods as short as 100 days, notably shorter than the minimum orbital periods of 600 days in the solar grid. 

The above discussion suggests that within the effective temperature range of $4000-8500$ K, post-RGB binaries are expected to dominate in their luminosity regime. We quantify this in Sect. \ref{probabilities}.

\subsection{Post-RGB luminosity-orbital period relations} \label{relation}
Fig. \ref{figure:models} shows that the post-RGB stars with initial primary masses below 2.0 $M_{\odot}$ exhibit a tight luminosity-orbital period relation. This stems from the fact that RGB stars follow a core mass-radius relation \citep[e.g.][]{Rappaport1995}. The occurrence of stable mass transfer results in the radius of the primary being approximately equal to its Roche lobe radius, which in turn is determined by the orbital separation of the binary \citep[as in Eq. 2 from][]{Eggleton1983}, and thus to the orbital period, using Kepler's third law. In addition, RGB and post-RGB stars follow a strict core mass-luminosity relation \citep[e.g.][]{Refsdal1970,Kippenhahn1981}, which results in the observed luminosity-orbital period relation.

Our post-RGB models reach maximum luminosities of about 2300 $L_\odot$ and 1800 $L_\odot$ for the solar and metal-poor grids, respectively, which is consistent with the expectation that they should be below the RGB-tip luminosity (2500 $L_\odot$). Although the exact value of the RGB-tip luminosity may vary, the value we use is consistent with our models as an upper limit.

\begin{figure}
    \resizebox{\hsize}{!}{\includegraphics{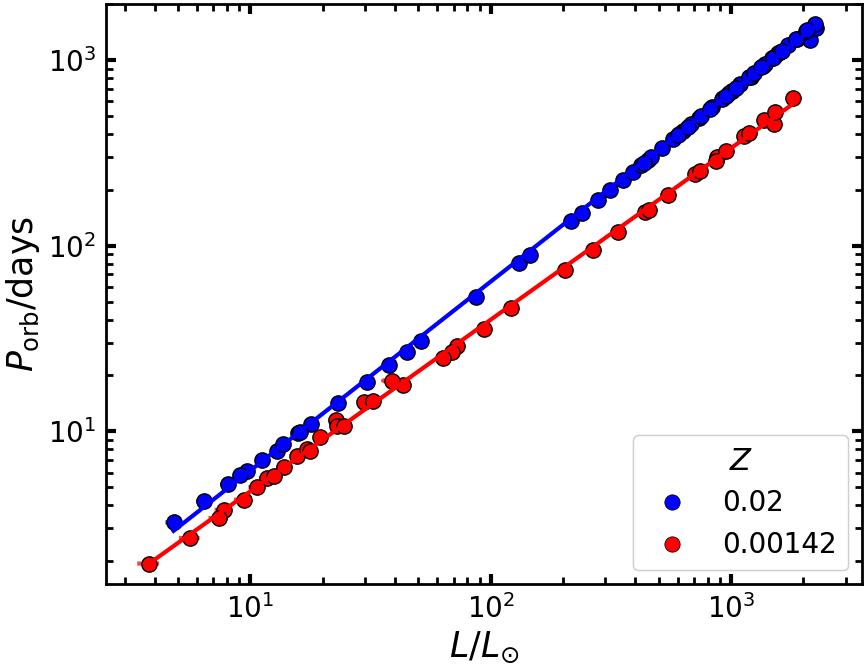}}
    \caption{Orbital period-luminosity plot of our post-RGB models with initial primary masses below $2.0$ $M_{\odot}$, indicated by markers and colour-coded by metallicity. Corresponding fits for the solar and metal-poor grids, given in Eq. \ref{equation:LPorb}, are denoted by solid lines.}
    \label{figure:LPorb_relations}
\end{figure}

The post-RGB luminosity-orbital period relations for the two metallicities are compared in Fig. \ref{figure:LPorb_relations}, which shows that the relation for $Z=0.00142$ lies at shorter orbital periods than the relation for $Z=0.02$. This results from the fact that lower metallicity RGB stars are more compact \citep[e.g.][]{Rappaport1995}. The orbital periods of binaries of comparable luminosity can vary by a factor of about two over the metallicity range of the Galactic sample. Our post-RGB models for the solar and metal-poor grids reach orbital periods of up to 1600 and 620 days, respectively.

The luminosity-orbital period relations exhibited by our models of post-RGB stars with initial primary masses less than 2.0 $M_{\odot}$ follow an almost power-law relationship, which can be well fitted by
\begin{equation}
    \log(P_\mathrm{orb}/\mathrm{days}) = a_0 + a_1 \log(L/L_{\odot}),
    \label{equation:LPorb}
\end{equation}
with the following coefficients for the two metallicities:
\begin{equation*}
Z=0.02:
\begin{cases}
    a_0 = -0.227\pm0.006, \\
    a_1 = 1.017\pm0.002, \\
\end{cases}
4.8<L/L_{\odot}<2300, \\
\end{equation*}
\begin{equation*}
Z=0.00142:
\begin{cases}
    a_0 = -0.246\pm0.009, \\
    a_1 = 0.924\pm0.004, \\
\end{cases}
3.8<L/L_{\odot}<1800. \\
\end{equation*}

\subsection{Observational probabilities} \label{probabilities}
We computed the probability to observe a certain mass-transfer product from our models as the product of their evolutionary timescales and their birth rates. We defined the evolutionary timescales as the elapsed time during which the models satisfy the selection criteria described in Sect. \ref{selection}. For the calculation of these observation probabilities, we considered a volume-limited sample and did not take observational bias into account.

We used the multidimensional binary parameter distributions from \citet{Moe2017} to determine the birth rates of our models. These distributions describe main-sequence binaries in terms of primary mass, orbital period, and mass ratio\footnote{We also apply a correction to the orbital parameters for the mass lost through stellar wind from the primary between the ZAMS and the onset of RLOF \citep{Soberman1997}, although this is insignificant in almost all cases.}. We computed the birth rates by integrating between the grid boundaries of each parameter (primary mass, orbital period, and mass ratio), using the initial binary sampler code described in \citet{Breivik2020} and changing the initial mass function to that given in \citet{Kroupa2001}. We defined the grid boundaries as the midpoints between the values of neighbouring models in the grid of the same metallicity. For the lower limit of the initial mass range, we used 0.9 $M_\odot$, which is approximately the minimum initial mass of a star that could have reached the post-main sequence. For the upper limit, we used 8.0 $M_\odot$. The mass ratios range from $0.1$ to $1.0$. Note that, since our model grids only sparsely cover the parameter space at discrete points, the values computed using this method are only rough estimates. 

In the full post-RGB luminosity range of our solar grid ($100-2300$ $L_\odot$), we find that 64$\%$ of objects are low-mass post-RGB stars, 23$\%$ are intermediate-mass post-RGB stars, 11$\%$ are CHeB stars, and 2$\%$ are post-AGB stars. For the metal-poor grid with luminosities of $100-1800$ $L_\odot$, we find that these values are 60$\%$, 20$\%$, 16$\%$, and 4$\%$, respectively. Lower luminosity post-RGB stars dominate these fractions, as they have much longer evolutionary timescales than their higher luminosity counterparts (see Appendix \ref{lifetimes} for more details). 

Our model grids predict that post-AGB stars overlap with post-RGB stars in the luminosity ranges of $600-2300$ $L_\odot$ and $600-1800$ $L_\odot$ for our solar and metal-poor grids, respectively. We calculate that in these ranges, the ratio of post-RGB to post-AGB stars is about 13:1 and 4:1 for the solar and metal-poor grids, respectively. Post-RGB stars are therefore also more abundant at these luminosities. A lower metallicity appears to increase the probability of observing a post-AGB star. To some extent, this may be due to the coarseness of the metal-poor grid. When we artificially increase the spacing in mass ratio for the solar grid to match that of the metal-poor grid (0.2 instead of 0.1), we find a ratio of post-RGB to post-AGB stars of 9:1 rather than 13:1. Nevertheless, this exercise suggests that the difference between the two metallicities is likely real. As these metal-poor post-AGB stars can have orbital periods up to five times shorter than post-RGB stars of comparable luminosity, as shown in Fig. \ref{figure:models}b, this may allow the two types of object to be distinguished.

All of our pre-subdwarf models have orbital periods close to the tip of the post-RGB luminosity-orbital period relation. We find that these pre-subdwarf models are between 20 to 200 times less likely to be observed than a post-RGB model at the tip of the relation. This is mainly because of their significantly shorter evolutionary timescales within the effective temperature range of $4000-8500$ K.

\section{Interpreting the observations} \label{comparison to data}
Our model grids predict that the majority of binaries with luminosities below 2500 $L_\odot$ contain post-RGB stars (see Sect. \ref{probabilities}). Moreover, these objects should follow the luminosity-orbital period relations presented in Sect. \ref{relation}. This means that, assuming the stable mass transfer formation channel, most of the observed objects in the samples with luminosities below 2300 $L_\odot$ should lie on these luminosity-orbital period relations.

\subsection{Sample versus luminosity-orbital period relations}
\begin{figure*}
\centering
   \includegraphics[width=18cm]{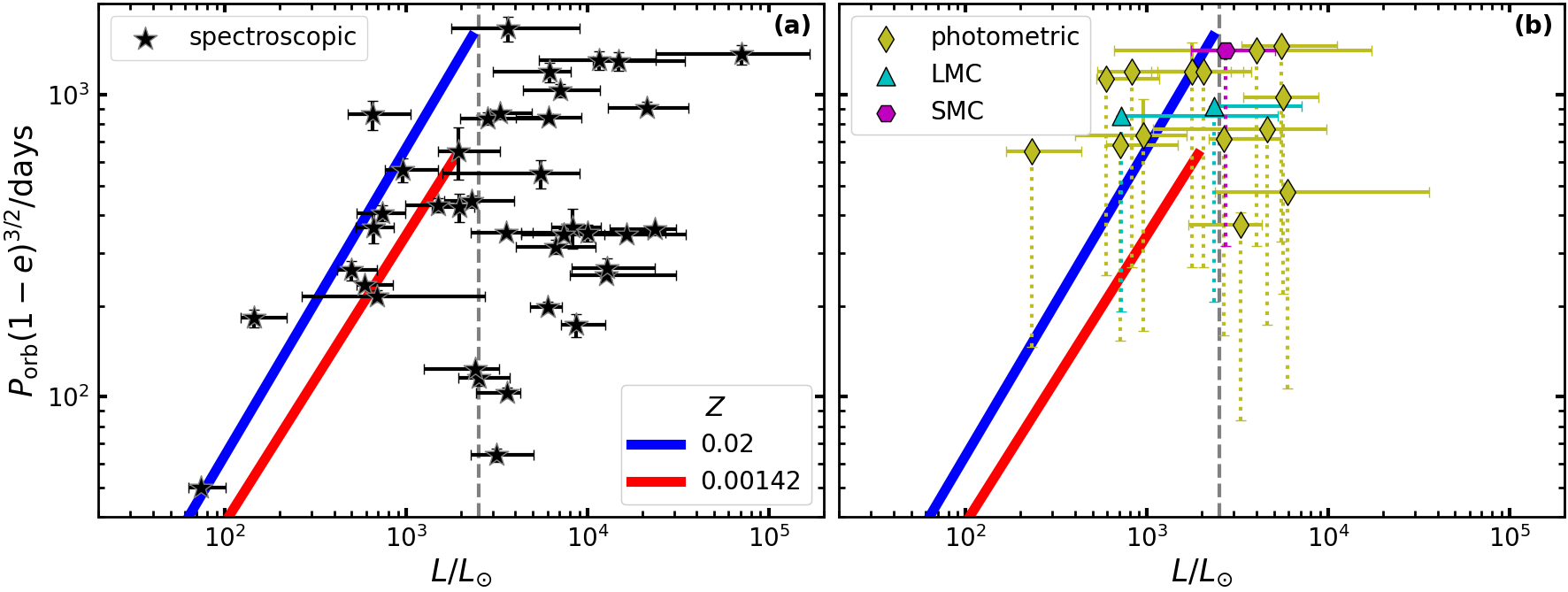}
     \caption{Orbital period-luminosity diagram of the post-AGB and post-RGB binary sample. Panel (a) shows Galactic objects with spectroscopically determined orbits. Panel (b) shows objects with photometrically determined orbits from the samples of the Milky Way, the LMC, and the SMC. The orbital periods are multiplied by a factor of $(1-e)^{3/2}$, which allows for the comparison of our models to systems with eccentric orbits, since in eccentric orbits mass transfer would occur during periastron \citep[see e.g.][]{Joss1987, Rappaport1995}. Solid error bars show observational uncertainties. The objects with photometrically determined orbital periods have unknown eccentricities, and their markers correspond to an assumed circular orbit. Dotted error bars represent the range in $P_\mathrm{orb}(1-e)^{3/2}$ assuming that their eccentricity falls between $0-0.63$, as observed for the sample with spectroscopic orbits. The solid lines show the luminosity-orbital period relations given by Eq. \ref{equation:LPorb}. The dashed grey lines correspond to the RGB-tip luminosity of 2500 $L_{\odot}$.}
     \label{figure:observations}
\end{figure*}

In Fig. \ref{figure:observations} we compare the Galactic sample with the theoretical post-RGB binary luminosity-orbital period relations. We focus the comparison on the 21 Galactic objects that fall within the post-RGB luminosity regime, since more luminous objects must contain post-AGB stars that we cannot fully interpret with our models. Of the post-RGB subsample, 14 objects have spectroscopic orbits and seven have photometrically determined orbital periods, shown in Figs. \ref{figure:observations}a and \ref{figure:observations}b, respectively. 

A direct comparison between the luminosity-orbital period relations and the post-RGB subsample is complicated by the fact that our models assume circular orbits, whereas many of the binaries in the sample are eccentric. Assuming that these eccentric binaries have undergone stable mass transfer, the point at which the donor last filled its Roche lobe at the end of mass transfer would be when the two binary components were closest to each other, i.e. at periastron. Therefore, in order to compare the orbital periods of these eccentric objects with our circular binary models, we converted the observed orbital periods to the orbital periods of a circular orbit with an orbital separation equal to the periastron distance by multiplying the orbital periods by a factor of ${(1-e)^{3/2}}$ \citep[see e.g.][]{Joss1987, Rappaport1995}. For the objects with photometrically determined orbital periods, the eccentricities are unknown. Therefore, in Fig. \ref{figure:observations}b we show the possible range in ${P_\mathrm{orb}(1-e)^{3/2}}$ for these objects, using an eccentricity range of $0-0.63$ as observed for objects with spectroscopically determined orbits. It is worth noting that binaries that have undergone stable RLOF are not a priori expected to have eccentric orbits (see Sect. \ref{implications} for a detailed discussion). 

Of the 14 objects with spectroscopically determined orbits shown in Fig. \ref{figure:observations}a, nine are consistent within their error bars with at least one of the luminosity-orbital period relations. Of the remaining five objects, four have luminosity errors within a factor of 1.3 of the relations. As these luminosity errors correspond to the $68\%$ confidence level (discussed further in Sect. \ref{observational limitations}), these numbers are consistent with the hypothesis that the sample is well described by the luminosity-orbital period relations. 

The remaining binary (BD+46~442) is located around 2400 $L_\odot$ and ${P_\mathrm{orb}(1-e)^{3/2}}$ of 120 days. Its luminosity is marginally consistent with a post-RGB star, but it has ${P_\mathrm{orb}(1-e)^{3/2}}$ about an order of magnitude shorter than expected from the relations. Since its luminosity is close to the RGB-tip luminosity, it could be a post-AGB star. We therefore exclude BD+46~442 from the post-RGB subsample. However, if this object turns out to be a post-RGB star, it cannot have been formed by stable mass transfer. 

There are seven objects with spectroscopic orbits above the RGB-tip luminosity that have lower luminosity limits below 2500 $L_\odot$, three of which are consistent with the post-RGB relations if their luminosities are in fact below the RGB-tip luminosity. Three other of these objects appear similar to BD+46~442 in that they have ${P_\mathrm{orb}(1-e)^{3/2}}$ that are too short to be consistent with the predicted post-RGB relations. They cannot have formed by stable mass transfer with an RGB donor and may contain a post-AGB star. 

The objects with photometrically determined orbital periods, shown in Fig. \ref{figure:observations}b, have considerable uncertainties on both luminosity and ${P_\mathrm{orb}(1-e)^{3/2}}$, making it difficult to draw concrete conclusions about the luminosity-orbital period relations. Nevertheless, within these uncertainties, all seven objects seem consistent with the relations. However, two of these objects, R~Sge and BT~Lac, would require eccentricities of at least 0.22 and 0.42, respectively, for their luminosity error bars to intersect the closest of the two relations. 

Of the five objects in Fig. \ref{figure:observations}b that are above the RGB-tip luminosity and have lower luminosity limits below 2500 $L_\odot$, three could be consistent with the post-RGB relations, assuming that their luminosity is indeed below the RGB-tip luminosity. In addition, the three objects in the Magellanic Clouds with photometric orbital periods, also shown in Fig. \ref{figure:observations}b, have luminosity errors that fall within the post-RGB luminosity regime and appear to be broadly consistent with the relations, although they have significant upper errors that may indicate that they contain post-AGB stars. 

The post-RGB subsample appears broadly consistent with the luminosity-orbital period relations. However, the uncertainties on the luminosities are large, preventing more stringent conclusions.

\subsection{Orbital period estimations of post-RGB binaries} \label{predicted periods}
There are many post-RGB objects in the Galactic, LMC, and SMC samples for which we lack orbital period determinations. Assuming that these have formed by stable mass transfer, we can use the luminosity-orbital period relations determined from our models to estimate their expected orbital periods.

In the Galactic sample, there are 17 objects with luminosities below the RGB-tip luminosity and without orbits. The depletion phenomenon makes it difficult to constrain the metallicity of the objects, so we used both the solar and metal-poor relations to estimate the orbital periods, which are presented in Table \ref{table:Galacticperiods}. For objects below the RGB-tip luminosity, but with luminosities above the range covered by the relations, the maximum luminosity of the relation was used instead. 

In the LMC and SMC samples, there are 56 and 19 post-RGB objects, respectively, for which we can estimate the orbital periods. Unlike for Galactic objects, the metallicity of these objects is well constrained by their association with the LMC or SMC, but we do not have binary models for the appropriate metallicities. We constructed luminosity-orbital period relations for the LMC and SMC using the RGB core mass-luminosity and core mass-radius relations from single-star models taken from the MIST database of the appropriate metallicity ($\mathrm{[Fe/H]}=-0.3$ and $-0.65$, respectively) and converting the radius to an orbital period by assuming stable mass transfer (see Appendix \ref{single star estimation} for a full explanation). The estimated orbital periods for the LMC and SMC are presented in Table \ref{table:MCsresults}. 

Future measurements of the orbital periods of these systems can provide a test of the proposed stable mass transfer formation channel for post-RGB binaries. This can be done by employing long-term radial velocity monitoring of such objects \citep[e.g.][]{Oomen2018}. The predicted orbital periods can be used to estimate the length of the observing campaign required for each possible post-RGB target.

\subsection{Post-RGB ages} \label{ages}
After mass transfer ends, post-RGB binaries may undergo a number of processes that modify their evolution and their orbits, depending on their post-mass transfer lifetimes. Such processes include eccentricity pumping via disc-binary interactions \citep[e.g.][]{Oomen2020}, depletion via re-accretion \citep[e.g.][]{Oomen2019}, and disc dissipation via slow outflows \citep[e.g.][]{Bujarrabal2016,GallardoCava2021,GallardoCava2023} and jet formation \citep[e.g.][]{Bollen2022,Verhamme2024,DePrins2024}. To understand the effect of these processes, evolutionary timescales need to be quantified. Therefore, we estimated the ages of the post-RGB stars in the samples by comparing their location in the HRD with our grids of post-RGB models with initial primary masses below 2.0 $M_{\odot}$. We defined this post-RGB age as the time elapsed since the end of mass transfer, taken as the moment when the mass transfer rate falls below $10^{-12}$ $M_{\odot}$/years. 

\begin{figure*}
\centering
   \includegraphics[width=18cm]{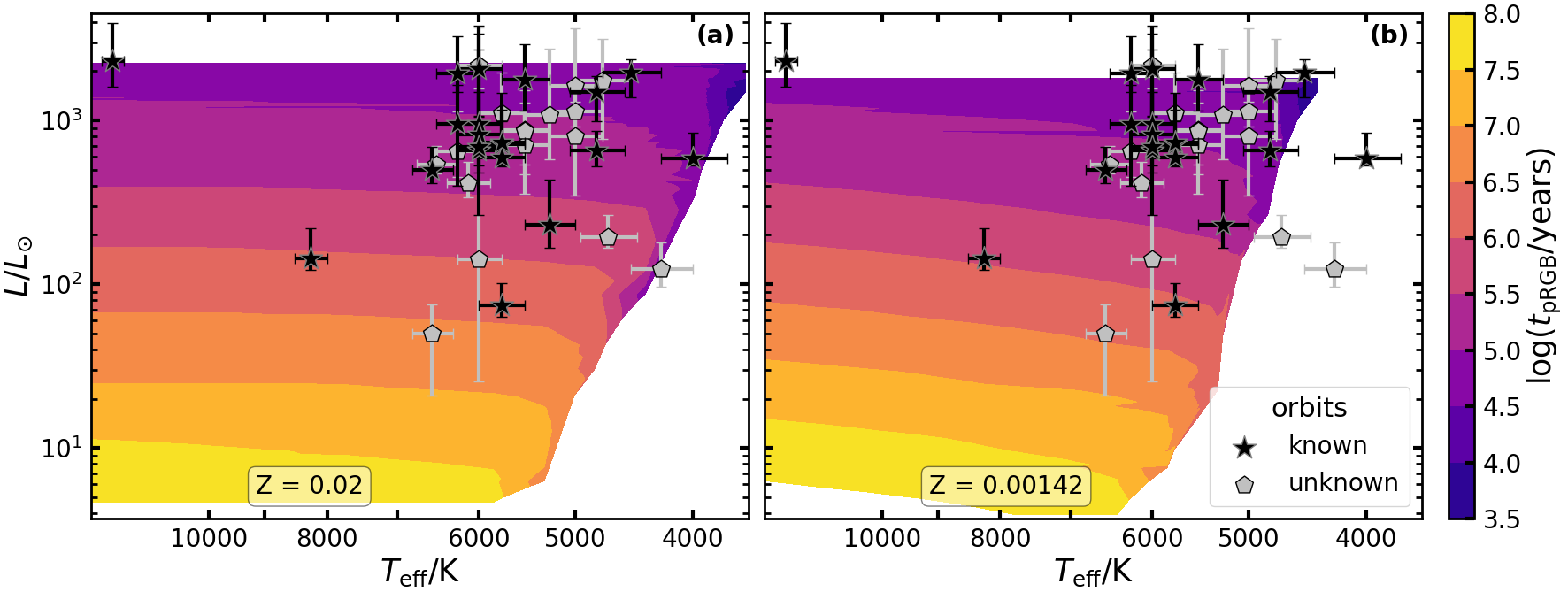}
     \caption{HRD of the Galactic sample with luminosities below the RGB-tip luminosity of 2500 $L_{\odot}$. Different markers denote whether the orbital periods of the objects are known. The colour scale indicates the elapsed time from the end of mass transfer in our post-RGB models, defined as the age of the post-RGB star, for the solar and metal-poor grids in panels (a) and (b), respectively.}
     \label{figure:ages}
\end{figure*}

In Fig. \ref{figure:ages} we show the HRD with a colour scale constructed from the evolutionary tracks of our post-RGB models, depicting the time elapsed since the end of mass transfer. The area covered by the colour scale is bound at high luminosity by the maximum luminosity of our post-RGB models and at low effective temperature by our definition of the end of mass transfer. Figs. \ref{figure:ages}a and \ref{figure:ages}b correspond to our solar metallicity and metal-poor models, respectively. A few models were removed when constructing the colour scale shown in Fig. \ref{figure:ages} because they deviated from the overall age trends of our post-RGB models (see Appendix \ref{lifetimes} for a full explanation). 

Post-RGB stars evolve to a higher effective temperature at approximately constant luminosity after mass transfer ends (see e.g. Fig. \ref{figure:HRD_Galactic}). Initially, the contraction is slow, so that the elapsed time increases quite rapidly as the effective temperature increases. After the effective temperature has increased by a few hundred K, the contraction accelerates and the lines of constant elapsed time become nearly horizontal in the HRD, as can be seen in Fig. \ref{figure:ages}. The elapsed time depends strongly on the luminosity, because a higher luminosity means a higher burning rate, which accelerates the decrease in the envelope mass, which in turn accelerates the increase in the effective temperature \citep{Chen2017}.

Comparing the two panels shown in Fig. \ref{figure:ages}, we see that lower metallicity models are expected to have lower post-RGB ages at low effective temperatures. This is because mass transfer ends at higher effective temperatures for such models due to their RGB donors having higher effective temperatures. This metallicity difference in post-RGB ages disappears at high effective temperatures.

In Fig. \ref{figure:ages}, we also plot the 37 objects from the Galactic sample that have luminosities below the RGB-tip luminosity of approximately 2500 $L_{\odot}$. By interpolating the post-RGB model grid of each metallicity, we estimated the ages of these objects, which are presented in Table \ref{table:Galacticages}. The errors were estimated as the largest and smallest ages that intersect with the error box. For objects with luminosities below 2500 $L_\odot$ but above the maximum luminosity of our grids, we used the maximum luminosity of the grid to estimate the post-RGB age. Objects that fall below the minimum effective temperature of our grids were assigned an age of zero, as these objects may still be undergoing mass transfer. The post-RGB ages determined in this way range between 0 and $4\times10^6$ years, with a median age of $10^5$ and $8\times10^4$ years for solar and metal-poor metallicity, respectively. 

Using this same method, we estimated the ages of the post-RGB objects in the LMC and SMC samples, which are presented in Table \ref{table:MCsresults}. The metallicities of the LMC and SMC are between those of our solar and metal-poor grids, which means that the post-RGB ages fall within the range predicted by our two grids. We find that the ages for the LMC range between 0 and $2\times10^6$ years, with a median age of $10^5$ years for both solar and metal-poor metallicity. For the SMC, the ages range between 0 and $5\times10^5$ years, with a median age of $2\times10^5$ and $10^5$ years for solar and metal-poor metallicity, respectively. 

\section{Discussion} \label{discussion}
\subsection{Observational limitations} \label{observational limitations}
The luminosities we determined for the sample objects have substantial uncertainties associated with them. These result from the errors on the distances of \citet{BailerJones2021} and on the reddening determinations we performed. 

The distances of \citet{BailerJones2021} may be inaccurate for the sample objects because the orbital motion of unresolved binaries can affect the parallax measurement of Gaia. However, this effect is only expected to be significant for objects with orbital periods close to 1.0 year, as it can mimic parallactic motion \citep{Penoyre2020}. Only three objects in the Galactic sample meet this criterion, all of which have luminosities above the RGB tip.

In addition, the reddening determinations are difficult to perform because many of the sample objects are photometrically variable, which means that the photometric measurements of an object are unlikely to all be in the same pulsation phase, so there may be significant noise in parts of their constructed SEDs \citep{Kluska2022}. Although we do not expect this to have a systematic effect, for a small subset of objects it could significantly affect their determined luminosities.

Furthermore, the uncertainties on the luminosities are not Gaussian. The errors provided by \citet{BailerJones2021} are the 16th and 84th percentiles of their posterior used as the distance estimate. Therefore, the error interval we provide corresponds to a confidence level of 68$\%$ and cannot be used to compute other confidence intervals.

A major observational limitation when investigating the formation of post-AGB and post-RGB binaries is that we lack the orbits of 32 objects in the Galactic sample, as well as the eccentricities of the 14 objects with photometrically determined orbital periods. Moreover, we lack the orbits of almost all objects in the LMC and SMC samples.

We expect Gaia DR4 to decrease the uncertainties on the luminosity determinations of the Galactic post-RGB and post-AGB binaries. This is because the parallaxes will improve by at least a factor of $\sqrt{2}$ compared to Gaia DR3, and this factor is expected to be even higher for these objects since they are all optically bright sources \citep{Brown2025}. Furthermore, Gaia DR4 may provide the missing orbital parameters of the 32 objects via improved astrometry or radial velocity data.

\subsection{Limitations of our models}
For all our models, we assumed conservative mass transfer, following \citet{Temmink2023}. However, the circumbinary discs observed around post-AGB and post-RGB binaries indicate that at least some mass loss from the system must have occurred. Moreover, our models of mass transfer from RGB and early-AGB donors show substantial peak mass-transfer rates \citep{Temmink2023}, which suggests that high mass-transfer efficiencies are unlikely due to the long thermal timescales of the main-sequence companions \citep[e.g.][]{Hurley2002}. In addition, there is evidence from other types of post-mass-transfer binaries to suggest that mass transfer involving an RGB donor is at least partly non-conservative \citep[e.g.][]{Vos2018,Sun2021,Nine2023,Parsons2023}.

Nevertheless, many previous studies have shown that the core mass-orbital period relation is hardly affected by the non-conservative nature of stable mass transfer \citep{Tauris1999, DeVito2012, Chen2013, Smedley2014, Chen2017, Ge2024}. This is because the final orbital period of the binary is determined by the maximum radius achieved by the donor, which in turn is determined by its core mass and is independent of mass loss or angular momentum loss from the system, and only weakly dependent on the companion mass (see Appendix \ref{single star estimation} for further details). Thus, mass loss and angular momentum loss have only little effect on the orbital properties of binaries that have undergone stable mass transfer. For the purposes of this work, the assumption of conservative mass transfer does not meaningfully alter the luminosity-orbital period relations found and the conclusions drawn.

The Galactic sample of post-AGB and post-RGB binaries contains objects with significant eccentricities, ranging up to 0.63. At present, the observed eccentricities of post-RGB binaries cannot be explained by the stable mass transfer formation channel presented in this work. The binaries are expected to be circular at the onset of mass transfer due to tidal interactions \citep{Hurley2002, Oomen2020}. Nevertheless, the post-RGB subsample appears to be consistent with the post-RGB luminosity-orbital period relations, if we assume that the orbit remained eccentric during mass transfer and that the donor star filled its Roche lobe at periastron. Although interactions between the binary and the circumbinary disc can pump up the eccentricity following mass transfer, models thus far have not produced eccentricities of up to 0.5, as are seen for the Galactic post-RGB binaries \citep{Vos2015, Oomen2020}. 

As discussed in \citet{Temmink2023}, our grid contains models in which mass transfer is stable according to the quasi-adiabatic criterion, but unstable according to the dynamical timescale criterion (see Sect. \ref{selection}). We expect such systems to have shorter final orbital periods, since the donor would significantly overfill its Roche lobe. We identify a subset in the Galactic sample with luminosities around the RGB tip, but with orbital periods that are too short according to the luminosity-orbital period relations. Exploring this quasi-adiabatically stable but dynamically unstable parameter space could help shed light on this subsample of binaries.

In addition to post-RGB binaries, the samples include post-AGB binaries. Analogous to the work in this paper for RGB donors, it would be interesting to investigate whether stable mass transfer models with AGB donors can explain the properties of Galactic post-AGB binaries. In order to do so, we would need to extend our grid to include donors ranging beyond the first TP on the AGB. Model calculations by \citet{Ge2020,Ge2020b} suggest that mass transfer involving TPAGB donors can be stable for a range of mass ratios, although they show that such systems are prone to exceeding their outer Roche lobes. However, these studies do not appear to model the TPs themselves, despite the fact that these pulses have a significant impact on the radius evolution of the donor and, consequently, on the stability of mass transfer.

\subsection{Implications of our results} \label{implications}
The objects in the observed samples are characterised by the presence of circumbinary discs, identified by the IR excess in their SEDs. We expect these discs to form from matter that has not been accreted by the companion during mass transfer. The discs are thought to dissipate through processes such as re-accretion of gas onto the binary companion \citep{Oomen2019}, slow outflows with velocities around 10 km/s \citep[e.g.][]{Bujarrabal2016,GallardoCava2021,GallardoCava2023}, and jet formation with velocities of the order of 100 km/s \citep[e.g.][]{Bollen2022,Verhamme2024,DePrins2024}. \citet{Bujarrabal2016} used the disc/outflow mass ratio, derived from model fitting of CO line emission data, to estimate the disc dissipation timescales of several post-AGB binaries, ranging from $5\times10^3$ to more than $2\times10^4$ years. \citet{Bollen2022} and \citet{DePrins2024} estimated disc dissipation timescales from the jet feeding rate, derived from model fitting of spectroscopic data of several post-AGB and post-RGB binaries, to be between $10^2$ and $10^5$ years, assuming a typical disc mass of $10^{-2}$ $M_\odot$ \citep[e.g.][]{Bujarrabal2016,GallardoCava2021,GallardoCava2023}.

The post-RGB ages computed in Sect. \ref{ages} can also be interpreted as the ages of the circumbinary discs, since they are defined as the elapsed time after the end of mass transfer, at which point we expect the discs to stop receiving additional material and to begin to dissipate. Because the objects in the samples exhibit clear disc signatures, we infer that this dissipation is ongoing. The median age of about $10^5$ years that we estimated for the post-RGB subsamples is at the upper limit of the disc dissipation timescales predicted in previous works and around a factor of ten higher than the typical disc dissipation timescale. In the Galactic subsample three objects have lower age limits greater than $10^5$ years; for the LMC and SMC subsamples this is the case for 12 and two objects, respectively. Thus, about 15$\%$ of the objects have disc ages longer than any predicted disc dissipation timescale for these types of objects. Assuming that the post-RGB binaries in these samples are formed by the proposed stable mass transfer formation channel, this seems to suggest that they are observed towards the end of their circumbinary disc evolution, which would imply that current methods do not identify post-RGB binaries at earlier times with potentially more massive discs. Furthermore, we also predict a population of post-RGB binaries in which the circumbinary disc has dissipated before their effective temperature reaches 8500 K. These binaries are likely to have lower luminosities because of their long evolutionary timescales that greatly exceed estimated disc dissipation timescales (see Appendix \ref{lifetimes}). 

For the estimated post-RGB ages of the sample stars, we did not take into account that matter can be re-accreted from the circumbinary disc onto the primary, for which there is evidence in the form of depletion of refractory elements \citep[][and references therein]{Oomen2019}. Circumbinary material accreted onto the post-AGB or post-RGB star increases its envelope mass, which is expected to increase the evolutionary timescale of a high-luminosity post-RGB star by a factor of two to three \citep{Oomen2019}. This means that the true disc age could be higher by a similar factor, especially if the object in question exhibits depletion. Taking this into account reinforces the discrepancy between the estimated ages and the predicted disc dissipation timescales.

The Galactic sample contains few objects with low luminosities, with nine out of 85 objects below 500 $L_\odot$ (about 10$\%$ of the sample). The LMC sample has 22 out of 101 objects below 500 $L_\odot$, and the SMC sample has seven out of 27 (in both cases about 25$\%$). However, as discussed in Appendix \ref{lifetimes}, the evolutionary timescales of post-RGB stars decrease with core mass and thus with luminosity, by a factor of at least ten for a post-RGB star with a luminosity of 2000 $L_\odot$ compared to one with a luminosity of 200 $L_\odot$. Therefore, we would expect to see many more low-luminosity ($L<500$ $L_\odot$) post-RGB stars. In particular, the Galactic sample appears to be incomplete at these low luminosities when compared with the LMC and SMC samples. This is further illustrated by the fact that the LMC and SMC are at distances of 50 and 62 kpc, respectively \citep{Walker2012,Pietrzynski2013,Graczyk2013}, compared to the most distant low-luminosity Galactic post-RGB binary, which is at a distance of only 4.2 kpc. One possible explanation is that the circumbinary discs of these low-luminosity post-RGB binaries dissipate faster than their evolutionary timescale, such that they lose the signature IR excess used in systematic searches that make up the bulk of the samples \citep{deRuyter2006, Gezer2015, Kamath2014, Kamath2015}.  

Moreover, our stable mass transfer models produce post-RGB stars with luminosities as low as 5 $L_\odot$, which is an order of magnitude lower than any object in the Galactic, LMC, or SMC sample. Based on their evolutionary timescales, we expect post-RGB stars with a luminosity of 5 $L_\odot$ to be 100 times more numerous than those with a luminosity of 200 $L_\odot$. However, AU~Peg is the lowest-luminosity post-RGB binary with a known orbit, with a luminosity of 74 $L_\odot$, which is consistent with the luminosity-orbital period relations. Despite the samples specifically selecting optically bright sources, the lack of these extremely low-luminosity ($L<50$ $L_\odot$) post-RGB binaries indicates that they may not form circumbinary discs, or at least not ones that produce a measurable IR excess in their SEDs. 

EL~CVn binaries are a similar class of object that has been observed, albeit with shorter orbital periods of $0.46-3.8$ days and lower luminosities of $0.32-8.4$ $L_\odot$ \citep{VanRoestel2018}. They consist of a low-mass helium white dwarf precursor with an effective temperature of $8000-17000$ K and a main-sequence companion, and are consistent with a stable mass transfer formation scenario \citep{Chen2017}. While these EL~CVn binaries confirm that stable mass transfer creates post-RGB objects with extremely low luminosities, there is still an apparent lack of objects with luminosities between 5 and 50 $L_\odot$. We predict the existence of a population of such post-RGB binaries that likely lacks the near-IR excess that is currently used for their identification.

\section{Conclusion} \label{conclusion}
We investigated whether the properties of the Galactic post-RGB binary sample could be explained by the stable mass transfer formation channel. We used a comprehensive grid of solar metallicity binary evolution models from \citet{Temmink2023} and computed a similar grid of metal-poor models. We selected the stable mass transfer models with post-mass-transfer primaries, whose effective temperatures fall within the observed range of post-RGB binaries ($4000-8500$ K). 

We find that our post-RGB models with initial primary masses below $2.0$ $M_{\odot}$ follow strict luminosity-orbital period relations. Although our models assumed conservative mass transfer, these relations are expected to be independent of mass-transfer efficiency. Among the different types of post-mass-transfer binaries from RGB and early-AGB donors with effective temperatures between 4000 and 8500 K that we identified in our grid, we find that low-mass post-RGB binaries are expected to be most numerous below the RGB-tip luminosity based on their birth rates and evolutionary timescales.

We conclude that the Galactic sample of post-RGB binaries with spectroscopically determined orbits appears consistent with the luminosity-orbital period relations. This is also broadly the case for the post-RGB binaries with photometrically determined orbital periods, although they have larger uncertainties because their eccentricities are unknown. At present, we cannot explain the observed eccentricities within the stable mass transfer formation channel. Tidal forces are expected to have circularised the orbit before the onset of RLOF, and no post-mass-transfer eccentricity-pumping mechanism has been found to be strong enough.

We computed the post-mass-transfer ages of post-RGB stars by comparing their location in the HRD with our post-RGB models. We estimate that the Galactic post-RGB subsample has a median age of $10^5$ years, which is about an order of magnitude longer than the median disc dissipation timescale presented in previous papers. In addition, 15$\%$ of the post-RGB binaries have lower age limits above the maximum estimated disc dissipation timescale. This suggests that post-AGB binaries are observed towards the end of their circumbinary disc evolution, and that younger post-RGB binaries with potentially more massive discs may be missing from the samples. We also find that, if we do not take into account the dissipation of the circumbinary disc, the number of post-RGB stars is expected to increase towards lower luminosity, which is not the case in any of the post-RGB samples. Therefore, either the circumbinary discs need to have dissipated much faster than the evolutionary timescale of these post-RGB stars, or the discs do not form at all. 

The errors on the luminosities of the Galactic post-RGB binaries are substantial, as both the reddening determination and the distance measurement have large uncertainties. This means that it is difficult to rigorously validate the stable mass transfer formation scenario when we compare the post-RGB luminosity-orbital period relations with the Galactic sample. While the LMC and SMC samples have smaller luminosity uncertainties due to their well-determined distances, we lack spectroscopically determined orbital parameters for these objects. With the help of long-term radial velocity monitoring of post-RGB stars in the LMC and SMC, the orbital properties combined with their known luminosities would allow for the post-RGB luminosity-orbital period relations to be tested more accurately. The estimated orbital periods of these objects, presented in Table \ref{table:MCsresults}, can be used to select the best possible candidates for such spectroscopic observations. In addition, we expect Gaia DR4 to improve the luminosity determinations of the Galactic post-RGB and post-AGB binaries due to more accurate parallax measurements. Furthermore, Gaia DR4 may provide the missing orbital parameters of many of these objects via improved astrometry or radial velocity data.

\section*{Data availability}
Tables \ref{table:GalacticSEDs}, \ref{table:MCsSEDs}, \ref{table:solarmodels}, \ref{table:mpmodels}, and \ref{table:MCsresults} are only fully available in electronic form at the CDS via \url{https://cdsarc.cds.unistra.fr/viz-bin/cat/J/A+A/703/A294}.

\begin{acknowledgements}
The authors thank the anonymous referee for their constructive comments. C.A.S.M. expresses gratitude to Mariya Nizovkina for invaluable input on improving the overall readability of the text and figures. K.D.T. acknowledges support from NOVA. This research made use of the SIMBAD database \citep{Wenger2000} and the VizieR catalogue access tool \citep{Ochsenbein2000} operated at the CDS, NASA's Astrophysics Data System, and the following Python software packages and tools: IPython \citep{Perez2007}, NumPy \citep{Harris2020}, Matplotlib \citep{Hunter2007}, and SciPy \citep{Virtanen2020}.
\end{acknowledgements}

\bibliographystyle{aa}
\bibliography{references}

\begin{appendix}
\section{Estimating post-RGB relations from single-star tracks} \label{single star estimation}
The luminosity-orbital period relations that we present in this paper are taken from our low-mass post-RGB binary evolution models. Instead of constructing a grid of detailed binary models to determine post-RGB relations for donors with altered input parameters (e.g. different metallicity or mixing assumptions), it is possible to approximate these relations using single-star models of those donors, as has been done in many previous works starting from \citet{Rappaport1995}. In this appendix, we demonstrate how post-RGB relations can be accurately estimated using single-star evolutionary tracks.

In order to construct the luminosity-orbital period relation, the core mass-luminosity and core mass-radius relations followed by the RGB donor are used to give a relation between luminosity and radius, and the orbital period is determined from the radius. Since stable mass transfer entails that the donor fills its Roche lobe, the maximum radius reached by the donor at the end of mass transfer determines the final orbital period of the binary. The Roche-lobe radius is related to the orbital separation of the binary according to \citet{Eggleton1983}: 
\begin{equation}
    R_\mathrm{L,d}/a \approx \frac{0.49q^{-2/3}}{0.6q^{-2/3}+\ln\left(1+q^{-1/3}\right)},
    \label{equation:Eggleton1983}
\end{equation}
where $q=M_\mathrm{a}/M_\mathrm{d}$. Similarly as in \citet{Verbunt1993}, combining Eq. \ref{equation:Eggleton1983} with Kepler's third law allows us to express the orbital period as
\begin{equation}
    P_\mathrm{orb} = \frac{2\pi F(q) R_\mathrm{L,d}^{3/2}}{0.343G^{1/2} M_\mathrm{d}^{1/2}},
    \label{equation:Porb}
\end{equation}
where $F(q)$ is the mass ratio dependence expressed by
\begin{equation}
    F(q) = \frac{q}{\left(1+q\right)^{1/2}}  \left[0.6q^{-2/3}+\ln\left(1+q^{-1/3}\right)\right]^{3/2}.
    \label{equation:qdependence}
\end{equation}
The mass ratio dependence of the orbital period is extremely weak, particularly for post-mass-transfer systems with $q>1.0$. This means that the mass of the binary companion has a negligible effect on the final orbital period if the mass transfer is stable, as can be seen in Fig. \ref{figure:LPorb_relations}, where the low-mass post-RGB models strictly follow the luminosity-orbital period relation. 

\begin{figure}
    \resizebox{\hsize}{!}{\includegraphics{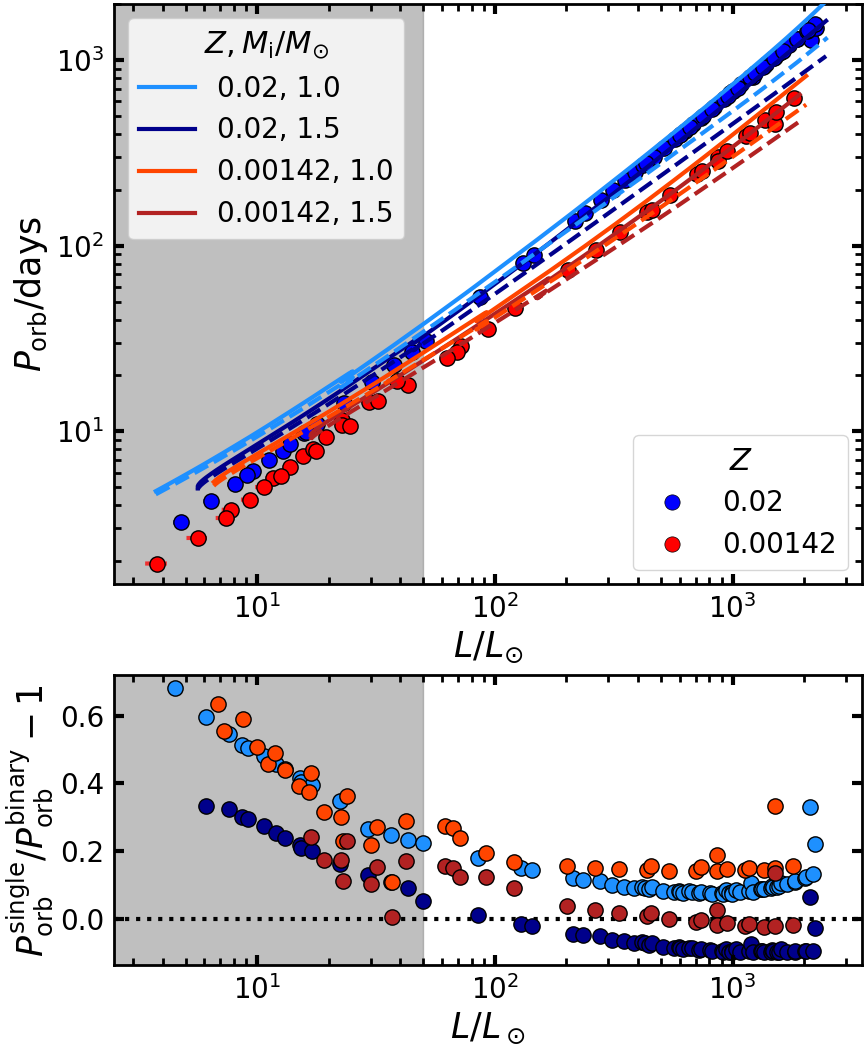}}
    \caption{Orbital period-luminosity plot of our binary post-RGB models with initial primary masses below $2.0$ $M_{\odot}$, indicated by markers and colour-coded by metallicity. Also shown are the luminosity-orbital period relations derived from single-star donor models, colour-coded by metallicity and initial mass. Solid and dashed lines show those relations with and without atmospheric RLOF, respectively. The observed luminosity range of post-RGB stars is denoted by the non-darkened area ($50<L/L_\odot<2500$). The bottom panel shows the relative difference between the estimated orbital period from the single-star models, $P_\mathrm{orb}^\mathrm{single}$, and the orbital period from the binary models, $P_\mathrm{orb}^\mathrm{binary}$, for the same luminosity, colour-coded by metallicity and initial mass of the single-star model.}
    \label{figure:LPorb_comparison}
\end{figure}

It is usually assumed that $R_\mathrm{L,d} = R_\mathrm{d}$ \citep[e.g.][]{Rappaport1995}. However, the donor radii in our binary evolution models are substantially smaller than their Roche lobes, as mass transfer occurs from the extended atmospheres of the donors \citep{Ritter1988,Pastetter1989}. Consequently, a star of a certain radius can fill its Roche lobe in a wider orbit. We find that atmospheric RLOF can be accurately approximated by
\begin{equation}
R_\mathrm{L,d} = R_\mathrm{d} + C H_\mathrm{P},
\label{equation:aRLOF}
\end{equation}
where $H_\mathrm{P}$ is the pressure scale height of the photosphere,
\begin{equation}
H_\mathrm{P}= \frac{k T_\mathrm{eff} R_\mathrm{d}^2}{\mu m_\mathrm{u} G M_\mathrm{d}},
\label{equation:HP}
\end{equation}
and $\mu$ is the mean molecular weight, which we approximate as 
\begin{equation}
\mu= (X + Y/4)^{-1},
\label{equation:mu}
\end{equation}
assuming that the outer envelope of the star is a neutral monatomic gas with the same composition as it had at the ZAMS and neglecting the metallicity contribution. For our solar and metal-poor grids, $(X,Y)$ are $(0.7,0.28)$ and $(0.74745,0.25113)$, respectively. We determined $C$ in Eq. \ref{equation:aRLOF} by fitting $R_\mathrm{L,d}-R_\mathrm{d}$ from our binary models versus $H_\mathrm{P}$ at the point where $R_\mathrm{d}$ is largest during the mass transfer episode. We find a tight linear relationship, with $C=9.19\pm0.03$ and $9.05\pm0.01$ for the solar and metal-poor grids, respectively, where the uncertainties correspond to the standard errors of these least-squares fitting parameters. For these fits, we excluded the outliers discussed in Appendix \ref{lifetimes}.  

The comparison of the single-star relations determined using the above-described method with the relations from our binary models is shown in Fig. \ref{figure:LPorb_comparison}. We used the single-star donor models with initial masses of $1.0$ and $1.5$ $M_\odot$ from the solar and metal-poor grids, and assumed that the companion had a mass of 0.8 $M_\odot$. For $M_\mathrm{d}$ in Eqs. \ref{equation:Porb} and \ref{equation:HP} we used the core mass of the single-star model, based on the assumption is that the star has reached the end of RLOF with its envelope stripped, leaving a negligible remaining envelope mass.

The binary models produce a slightly steeper luminosity-orbital period relation than the single-star models. This is caused by the smaller envelope mass in the binary models, which leads to a decrease in the luminosity and radius of the stripped star compared to a single star with the same core mass \citep{Refsdal1970}. The single-star relations for donors with higher initial masses predict systematically lower orbital periods because the core mass-radius relation depends weakly on the stellar mass \citep[e.g.][]{Joss1987}. The effect of atmospheric RLOF on the single-star relations is seen in Fig. \ref{figure:LPorb_comparison} by comparing the solid and dashed lines. Including it prevents the systematic underestimation of the orbital periods at high luminosities by up to 40$\%$. The bottom panel of Fig. \ref{figure:LPorb_comparison} shows that the single-star relations derived from $1.5$ $M_\odot$ donor models reproduce the orbital periods of the binary models within 16$\%$ for luminosities in the observed post-RGB luminosity range, and within 10$\%$ for luminosities greater than 100 $L_\odot$. However, at luminosities below 50 $L_\odot$, the same relations systematically overestimate the periods by up to 33$\%$. Using a 1.0 $M_\odot$ rather than a 1.5 $M_\odot$ single-star model leads to a systematic overestimation of the orbital period by about 10$\%$ or more across the observed post-RGB luminosity range. Hence, $1.5$ $M_\odot$ donor models provide more accurate orbital period estimations. Note that the binary post-RGB models do not exhibit any such dependence of the orbital period on the initial donor mass.

We used the method described above to construct the luminosity-orbital period relations for the LMC and SMC, employing the luminosities and radii during the RGB phase of single-star MIST models with an initial mass of $1.5$ $M_\odot$, no initial rotation, and the appropriate metallicity (i.e. $\mathrm{[Fe/H]}=-0.3$ and $-0.65$, respectively). Since the value of $C$ in Eq. \ref{equation:aRLOF} is almost independent of metallicity, we used the mean of the values found from our model grids, which is $9.12$. The lowest-luminosity post-RGB star in the LMC and SMC samples has a luminosity of 80 $L_\odot$, for which the predicted overestimation of the orbital periods is only about 12$\%$. Therefore, this method should accurately estimate the expected orbital periods of the LMC and SMC samples.

\section{Post-RGB evolutionary timescales} \label{lifetimes}
\begin{figure}
    \resizebox{\hsize}{!}{\includegraphics{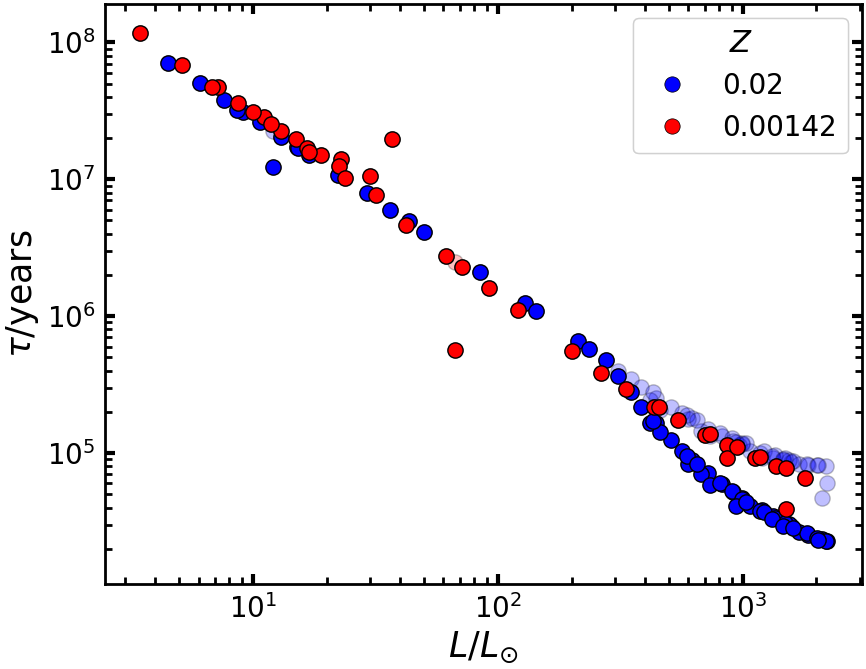}}
    \caption{Evolutionary timescale-luminosity plot of our post-RGB models with initial primary masses below $2.0$ $M_{\odot}$, indicated by markers and colour-coded by metallicity. Opaque markers show the evolutionary timescale from an effective temperature of 4000 K to 8500 K, while translucent markers show the evolutionary timescale from the end of mass transfer to an effective temperature of 8500 K.}
    \label{figure:Lduration_relations}
\end{figure}

Along with their luminosity and orbital period, an important property of post-RGB stars is their evolutionary timescale, i.e. the length of time they remain in the region of the HRD in which they are observed. Fig. \ref{figure:Lduration_relations} shows the luminosity-evolutionary timescale relations of our post-RGB models, where the evolutionary timescale is defined as the time that the post-RGB model spends with effective temperatures within the observed range of $4000-8500$ K. These evolutionary timescales do not take into account the effect of re-accretion, as observed via the depletion process \citep[][and references therein]{Oomen2019}.

There is an anti-correlation between the luminosity and the evolutionary timescale of a post-RGB star, with the evolutionary timescales of our post-RGB models decreasing from $10^8$ to $10^5$ years as the luminosity increases from 5 to 2300 $L_\odot$. Such a relation has also been found for other helium white dwarf precursors \citep{Istrate2014,Chen2017}. This relation arises because H-shell burning sustains the phase of constant luminosity exhibited by post-RGB stars. Consequently, the evolutionary timescale is proportional to the envelope mass divided by the rate of H-burning, which in turn is proportional to the luminosity of the star. The envelope mass that remains after mass transfer is set by a critical value, below which the donor ceases to fill its Roche lobe. Since the mass of the burning shell decreases steeply with increasing core mass, this critical value also decreases with increasing mass of the stripped donor star \citep{Refsdal1970}. A larger remaining donor mass results in a higher luminosity, which explains the anti-correlation. The luminosity-evolutionary timescale relations are almost independent of metallicity, as the increases in envelope mass and H-burning rate at lower metallicities cancel each other out \citep[see discussion in][]{Chen2017}. 

There are a few outliers visible in Fig. \ref{figure:Lduration_relations}. Some models have shorter evolutionary timescales than the general trend because mass transfer has removed more envelope mass than the critical value at which the star stops filling its Roche lobe. Mass transfer in these models is closer to the instability limit (see Sect. \ref{selection}) than in other models, which means that the donors overfill their Roche lobes to a greater extent, thereby enabling more of their envelopes to be stripped. For the metal-poor outlier with a luminosity of about 30 $L_\odot$, the evolutionary timescale is longer because the envelope mass is larger than its critical value. This is related to the interruption of mass transfer when the burning shell reaches the H-discontinuity, which causes the radius to temporarily decrease. Mass transfer will not resume if this occurs when the donor has a sufficiently small envelope mass, resulting in a larger envelope mass at the end of mass transfer \citep{Istrate2016}.

The evolutionary timescales we derived from the models depend on the selected effective temperature window. If we choose the end of mass transfer as the lower limit of this window instead of an effective temperature of 4000 K, the evolutionary timescales of post-RGB models with luminosities above 300 $L_\odot$ in our solar grid are significantly longer, by up to a factor of six, as illustrated by the translucent markers in Fig. \ref{figure:Lduration_relations}. As these models have effective temperatures below 4000 K when mass transfer ends, they have burned a significant portion of their remaining envelope by the time they reach 4000 K.

\section{Supplementary tables} 

\begin{table*}
\caption{Orbital properties of the Galactic post-AGB and post-RGB binary sample.}
\label{table:orbits}
\centering
\begin{tabular}{c c c c c}
\hline\hline
IRAS & Name & $P_\mathrm{orb}/\mathrm{days}$ & $e$ & References \\
\hline
\multicolumn{5}{c}{Spectroscopically determined orbits} \\
\hline
21216+1803 & AU Peg & 53.3344$\pm$0.0003 & 0.043$\pm$0.003 & (1) \\
07140-2321 & SAO 173329; V421 CMa & 115.951$\pm$0.002 & 0$\pm$0.04 & (2) \\
19157-0247 & V1801 Aql & 119.6$\pm$0.1 & 0.34$\pm$0.04 & (2) \\
19135+3937 & V677 Lyr & 126.97$\pm$0.08 & 0.13$\pm$0.03 & (2) \\
01427+4633 & BD+46 442 & 140.82$\pm$0.02 & 0.085$\pm$0.005 & (2) \\
06452-3456 & ... & 215.4$\pm$0.4 & 0$\pm$0.03 & (2) \\
05208-2035 & AY Lep & 234.38$\pm$0.04 & 0$\pm$0.02 & (2) \\
22327-1731 & HD 213985; HM Aqr & 259.6$\pm$0.7 & 0.21$\pm$0.05 & (2) \\
06165+3158 & ... & 262.6$\pm$0.7 & 0$\pm$0.05 & (2) \\
17534+2603 & 89 Her; V441 Her & 289.1$\pm$0.2 & 0.29$\pm$0.07 & (2) \\
06176-1036 & HD 44179; V777 Mon & 317.6$\pm$1.1 & 0.27$\pm$0.03 & (2) \\
17279-1119 & HD 158616; V340 Ser & 363.3$\pm$1 & 0$\pm$0.1 & (2) \\
15469-5311 & ... & 390.2$\pm$0.7 & 0.08$\pm$0.02 & (2) \\
06472-3713 & ST Pup & 406$\pm$2.2 & 0$\pm$0.04 & (2) \\
10158-2844 & HR 4049; AG Ant & 430.6$\pm$0.1 & 0.3$\pm$0.01 & (2) \\
08544-4431 & V390 Vel & 501.1$\pm$1 & 0.2$\pm$0.02 & (2) \\
19125+0343 & BD+03 3950 & 519.7$\pm$0.7 & 0.24$\pm$0.03 & (2) \\
F15240+1452 & HD 137569 & 529.8$\pm$1 & 0.11$\pm$0.03 & (3) \\
12185-4856 & SX Cen & 564.3$\pm$7.6 & 0$\pm$0.06 & (2) \\
10456-5712 & HD 93662 & 572$\pm$6 & 0.4$\pm$0.07 & (4) \\
06338+5333 & HD 46703; V382 Aur & 597.4$\pm$0.2 & 0.3$\pm$0.02 & (2) \\
16230-3410 & V1333 Sco & 649.8$\pm$3.5 & 0$\pm$0.13 & (2) \\
04166+5719 & TW Cam & 662.2$\pm$5.3 & 0.25$\pm$0.04 & (2) \\
19472+4254 & DF Cyg & 784$\pm$16 & 0.4$\pm$0.1 & (5) \\
... & BD+39 4926 & 871.7$\pm$0.4 & 0.024$\pm$0.006 & (2) \\
12222-4652 & HD 108015; V1123 Cen & 906.3$\pm$5.9 & 0$\pm$0.03 & (2) \\
19163+2745 & EP Lyr & 1151$\pm$14 & 0.39$\pm$0.09 & (2) \\
18281+2149 & AC Her & 1188.9$\pm$1.2 & 0$\pm$0.05 & (2) \\
04440+2605 & RV Tau & 1198$\pm$17 & 0.5$\pm$0.1 & (5) \\
06034+1354 & DY Ori & 1248$\pm$36 & 0.22$\pm$0.08 & (2) \\
07008+1050 & HD 52961; PS Gem & 1288.6$\pm$0.3 & 0.23$\pm$0.01 & (2) \\
17038-4815 & V729 Ara & 1394$\pm$12 & 0.63$\pm$0.06 & (2) \\
14524-6838 & HD 131356; EN TrA & 1488$\pm$8.7 & 0.32$\pm$0.04 & (2) \\
12067-4508 & RU Cen & 1489$\pm$10 & 0.62$\pm$0.07 & (2) \\
09144-4933 & ... & 1762$\pm$27 & 0.3$\pm$0.04 & (2) \\
19199+3950 & HP Lyr & 1818$\pm$80 & 0.2$\pm$0.04 & (2) \\
11000-6153 & HD 95767; V802 Car & 1989$\pm$61 & 0.25$\pm$0.05 & (2) \\
07284-0940 & U Mon & 2550$\pm$143 & 0.25$\pm$0.06 & (2) \\
08005-2356 & V510 Pup & 2654$\pm$124 & 0.36$\pm$0.05 & (6) \\
\hline
\multicolumn{5}{c}{Photometrically determined orbital periods} \\
\hline
09060-2807 & BZ Pyx & 372$\pm$35 & ... & (7) \\
17243-4348 & LR Sco & 475 & ... & (8) \\
22223+5556 & BT Lac & 650 & ... & (9)(10) \\
06108+2743 & SU Gem & 682$\pm$2 & ... & (11) \\
17111-1819 & HD 155720 & 712 & ... & (8) \\
17233-4330 & V1504 Sco & 735$\pm$230 & ... & (7) \\
13085-6747 & PX Mus & 770 & ... & (8) \\
17530-3348 & AI Sco & 977$\pm$7 & ... & (11) \\
20117+1634 & R Sge & 1125 & ... & (10) \\
09538-7622 & GP Cha & 1190$\pm$300 & ... & (7) \\
19548+1951 & RS Sge & 1193.49$\pm$0.9 & ... & (12) \\
08011-3627 & AR Pup & 1194$\pm$4 & ... & (11) \\
20056+1834 & QY Sge & 1405.25$\pm$18.25 & ... & (13) \\
09256-6324 & IW Car & 1449$\pm$9 & ... & (11) \\
\hline
\end{tabular}
\tablefoot{The table lists the following information for each object: its IRAS name, alternative identifications, orbital period $P_\mathrm{orb}$, eccentricity $e$, and the reference(s) to the paper(s) from which these values were taken.}
\tablebib{(1) \citet{Csornyei2019}; (2) \citet{Oomen2018}; (3) \citet{Bolton1980}; (4) \citet{Maas2003}; (5) \citet{Manick2019}; (6) \citet{Manick2021}; (7) \citet{Kiss2007}; (8) \citet{Kluska2022}; (9) \citet{Bodi2019}; (10) \citet{Percy2015}; (11) \citet{Kiss2017}; (12)  \citet{Horne2012}; (13) \citet{Hrivnak2024}.}
\end{table*}

\begin{table*}
\renewcommand{\arraystretch}{1.25}
\caption{SED fitting results of the Galactic post-AGB and post-RGB binary sample.}
\label{table:GalacticSEDs}
\centering
\begin{tabular}{c c c c c c c}
\hline\hline
IRAS & Name & $T_\mathrm{eff}/\mathrm{K}$ & $D/\mathrm{pc}$ & $E(B-V)$ & $L/L_\odot$ & $L_\mathrm{IR}/L_*$ ($\%$) \\
\hline
... & EZ Gem & 6555 & 4216$_{-521}^{+630}$ & 0.43$_{-0.27}^{+0.04}$ & 50$_{-29}^{+25}$ & 53 \\
13258-8103 & ... & 6000 & 1608$_{-385}^{+704}$ & 1.58$_{-0.7}^{+0.39}$ & 50$_{-41}^{+251}$ & 284 \\
21216+1803 & AU Peg & 5750 & 597$_{-6}^{+6}$ & 0.08$_{-0.08}^{+0.14}$ & 74$_{-11}^{+27}$ & 19 \\
06108+2743 & SU Gem & 5750 & 1703$_{-146}^{+238}$ & 0$_{-0}^{+0.22}$ & 90$_{-15}^{+97}$ & 792 \\
08011-3627 & AR Pup & 6000 & 661$_{-63}^{+82}$ & 0.15$_{-0.15}^{+0.56}$ & 90$_{-32}^{+322}$ & 915 \\
20131+2554 & EF Vul & 4250 & 1790$_{-51}^{+60}$ & 0.16$_{-0.16}^{+0.22}$ & 124$_{-27}^{+55}$ & 35 \\
22327-1731 & HD 213985; HM Aqr & 8250 & 702$_{-22}^{+28}$ & 0.11$_{-0.04}^{+0.11}$ & 144$_{-22}^{+76}$ & 45 \\
06489-0118 & SZ Mon & 4700 & 1437$_{-103}^{+118}$ & 0.01$_{-0.01}^{+0.09}$ & 193$_{-27}^{+72}$ & 38 \\
17233-4330 & V1504 Sco & 6250 & 2485$_{-359}^{+495}$ & 0.25$_{-0.25}^{+0.08}$ & 203$_{-118}^{+148}$ & 469 \\
(...) & (...) & (...) & (...) & (...) & (...) & (...) \\
\hline
\end{tabular}
\tablefoot{The table lists the following information for each object: its IRAS name, alternative identifications, effective temperature $T_\mathrm{eff}$, distance $D$, reddening $E(B-V)$, luminosity $L$, and IR-to-stellar luminosity ratio $L_\mathrm{IR}/L_*$. The effective temperatures are taken from \citet{Kluska2022} and have a formal error of $\pm250$ K. The distances correspond to the geometric distances from \citet{BailerJones2021}. The whole table is available at CDS.
\tablefoottext{a}{Luminosities are taken from \citet{Menshchikov2002}, as no distance is available from \citet{BailerJones2021} due to the lack of a Gaia parallax measurement.}
}
\end{table*}

\begin{table*}
\renewcommand{\arraystretch}{1.25}
\caption{SED fitting results of the post-AGB and post-RGB binary candidate samples in the LMC and SMC.}
\label{table:MCsSEDs}
\centering
\begin{tabular}{c c c c c c c}
\hline\hline
Object & Host & $T_\mathrm{eff}/\mathrm{K}$ & $E(B-V)$ & $L/L_\odot$ & $L_\mathrm{IR}/L_*$ ($\%$) & Reference \\
\hline
J045745.46-683724.1 & LMC & 7131 & 0.0$_{-0.00}^{+0.28}$ & 81$_{-1}^{+115}$ & 67 & (2) \\
J054144.88-712351.7 & LMC & 4946 & 0.0$_{-0.00}^{+0.34}$ & 88$_{-0}^{+72}$ & 38 & (2) \\
J053811.63-680824.9 & LMC & 5421 & 0.11$_{-0.11}^{+0.13}$ & 94$_{-16}^{+28}$ & 27 & (2) \\
J053030.78-701805.5 & LMC & 4754 & 0.03$_{-0.03}^{+0.19}$ & 99$_{-4}^{+34}$ & 85 & (2) \\
J051412.13-702026.0 & LMC & 7133 & 0.05$_{-0.05}^{+0.27}$ & 112$_{-14}^{+149}$ & 92 & (2) \\
J045836.92-701120.1 & LMC & 8250 & 0.14$_{-0.14}^{+0.24}$ & 115$_{-39}^{+177}$ & 239 & (2) \\
J051751.41-713507.3 & LMC & 8225 & 0.28$_{-0.21}^{+0.17}$ & 121$_{-56}^{+102}$ & 72 & (2) \\
J045736.84-705127.2 & LMC & 10480 & 0.2$_{-0.07}^{+0.04}$ & 131$_{-35}^{+30}$ & 28 & (2) \\
J052107.90-714104.4 & LMC & 10500 & 0.13$_{-0.13}^{+0.10}$ & 131$_{-54}^{+86}$ & 81 & (2) \\
(...) & (...) & (...) & (...) & (...) & (...) & (...) \\
\hline
\end{tabular}
\tablefoot{The table lists the following information for each object: its identification, host Magellanic Cloud, effective temperature $T_\mathrm{eff}$, reddening $E(B-V)$, luminosity $L$, IR-to-stellar luminosity ratio $L_\mathrm{IR}/L_*$, and the reference to the paper from which the object was taken. The effective temperatures have a formal error of $\pm250$ K. The distances used to determine the luminosities were 50 kpc for the LMC \citep{Walker2012,Pietrzynski2013} and 62 kpc for the SMC \citep{Graczyk2013}. The whole table is available at CDS.}
\tablebib{(1) \citet{VanAarle2011}; (2) \citet{Kamath2015}; (3) \citet{Manick2018}; (4) \citet{Kamath2014}.}
\end{table*}

\begin{table*}
\caption{Initial parameters of utilised stable mass transfer models with $Z=0.02$.}
\label{table:solarmodels}
\centering
\begin{tabular}{c c c c}
\hline\hline
$M_\mathrm{i}/M_\odot$ & $M_\mathrm{RLOF}/M_\odot$ & $\log(R_\mathrm{RLOF}/R_\odot)$ & $q_\mathrm{min}$ \\
\hline
1.0 & 0.9996 & 1.3192 & 0.5 \\
1.0 & 0.9994 & 1.7642 & 0.8 \\
1.0 & 0.9988 & 6.8421 & 1.0 \\
1.0 & 0.9981 & 10.6907 & 0.9 \\
1.0 & 0.9972 & 16.8649 & 0.9 \\
1.0 & 0.9953 & 26.4747 & 0.8 \\
1.0 & 0.9908 & 41.5789 & 0.7 \\
1.0 & 0.9882 & 65.3414 & 0.7 \\
1.0 & 0.9788 & 102.7576 & 0.7 \\
(...) & (...) & (...) & (...) \\
\hline
\end{tabular}
\tablefoot{The table lists the following information for each primary model used to create the binary grid with $Z=0.02$: its initial mass $M_\mathrm{i}$, mass at the onset of RLOF $M_\mathrm{RLOF}$, radius at the onset of RLOF $R_\mathrm{RLOF}$, and the smallest mass ratio $q_\mathrm{min}$ in the model grid for this primary model to result in stable mass transfer. Note that the mass ratios were sampled in steps of $0.1$. There is a difference between $q_\mathrm{min}$ defined here and $q_\mathrm{crit}$ given in \citet{Temmink2023} for these same models: $q_\mathrm{min}$ indicates the range of mass ratios used in this work for each primary model, while $q_\mathrm{crit}$ gives an exact stability boundary in terms of mass ratio for that primary model. The whole table is available at CDS.}
\end{table*}

\begin{table*}
\caption{Initial parameters of utilised stable mass transfer models with $Z=0.00142$.}
\label{table:mpmodels}
\centering
\begin{tabular}{c c c c}
\hline\hline
$M_\mathrm{i}/M_\odot$ & $M_\mathrm{RLOF}/M_\odot$ & $\log(R_\mathrm{RLOF}/R_\odot)$ & $q_\mathrm{min}$ \\
\hline
1.0 & 0.9997 & 1.1475 & 0.4 \\
1.0 & 0.9995 & 1.3641 & 0.4 \\
1.0 & 0.9993 & 1.7192 & 0.4 \\
1.0 & 0.9992 & 2.0453 & 0.4 \\
1.0 & 0.9992 & 2.5743 & 1.0 \\
1.0 & 0.999 & 4.0722 & 1.0 \\
1.0 & 0.9987 & 6.0896 & 1.0 \\
1.0 & 0.9982 & 9.1175 & 1.0 \\
1.0 & 0.9973 & 14.4358 & 0.8 \\
(...) & (...) & (...) & (...) \\
\hline
\end{tabular}
\tablefoot{The table lists the following information for each primary model used to create the binary grid with $Z=0.00142$: its initial mass $M_\mathrm{i}$, mass at the onset of RLOF $M_\mathrm{RLOF}$, radius at the onset of RLOF $R_\mathrm{RLOF}$, and the smallest mass ratio $q_\mathrm{min}$ in the model grid for this primary model to result in stable mass transfer. Note that the mass ratios were sampled in steps of $0.2$. The whole table is available at CDS.}
\end{table*}

\begin{table*}
\caption{Estimated orbital periods of the Galactic post-RGB binary subsample without observed orbital periods.}
\label{table:Galacticperiods}
\centering
\begin{tabular}{c c c c c c c}
\hline\hline
Name & $P_\mathrm{orb}/\mathrm{days}$ & $P_\mathrm{orb}^\mathrm{min}/\mathrm{days}$ & $P_\mathrm{orb}^\mathrm{max}/\mathrm{days}$ & $P_\mathrm{orb}/\mathrm{days}$ & $P_\mathrm{orb}^\mathrm{min}/\mathrm{days}$ & $P_\mathrm{orb}^\mathrm{max}/\mathrm{days}$ \\
& $Z=0.02$ & $Z=0.02$ & $Z=0.02$ & $Z=0.00142$ & $Z=0.00142$ & $Z=0.00142$ \\
\hline
EZ Gem & 32 & 13 & 48 & 20 & 9 & 29 \\
EF Vul & 80 & 62 & 116 & 47 & 37 & 67 \\
IRAS 13258-8103 & 92 & 16 & 568 & 54 & 10 & 296 \\
SZ Mon & 125 & 107 & 173 & 72 & 62 & 97 \\
IRAS 22251+5406 & 273 & 221 & 368 & 149 & 122 & 197 \\
HD 186438 & 356 & 313 & 463 & 191 & 169 & 244 \\
CC Lyr & 428 & 297 & 757 & 227 & 161 & 387 \\
UY Ara & 469 & 354 & 1029 & 247 & 190 & 516 \\
HD 340949 & 529 & 227 & 872 & 276 & 125 & 442 \\
EZ Aql & 571 & 234 & 863 & 297 & 129 & 437 \\
GK Car & 581 & 310 & 1078 & 302 & 168 & 538 \\
V Vul & 720 & 382 & 1556\tablefootmark{a} & 369 & 204 & 633\tablefootmark{a} \\
IRAS 20094+3721 & 739 & 499 & 1323 & 378 & 262 & 633\tablefootmark{a} \\
V1711 Sgr & 754 & 611 & 1155 & 385 & 317 & 574 \\
CN Cen & 1097 & 504 & 1556\tablefootmark{a} & 548 & 264 & 633\tablefootmark{a} \\
GZ Nor & 1175 & 510 & 1556\tablefootmark{a} & 584 & 267 & 633\tablefootmark{a} \\
HD 114855; V956 Cen & 1456 & 1049 & 1556\tablefootmark{a} & 633\tablefootmark{b} & 525 & ... \\
\hline
\end{tabular}
\tablefoot{The table lists the following information for each object: its name and estimated orbital period $P_\mathrm{orb}$ determined from its luminosity for both metallicities (see Sect. \ref{relation}). $P_\mathrm{orb}^\mathrm{min}$ and $P_\mathrm{orb}^\mathrm{max}$ correspond to the orbital periods estimated using the luminosity upper and lower limits, respectively.
\tablefoottext{a}{Lower limit on the maximum orbital period, as the luminosity upper limit of the object is larger than the maximum luminosity of the binary models.}
\tablefoottext{b}{Lower limit on the orbital period, as the luminosity of the object is larger than the maximum luminosity of the binary models.}
}
\end{table*}

\begin{sidewaystable*}
\caption{Estimated orbital periods and post-mass-transfer ages of the post-RGB binary candidate subsamples in the LMC and SMC.}
\label{table:MCsresults}
\centering
\begin{tabular}{c c c c c c c c c c c}
\hline\hline
Object & Host & $P_\mathrm{orb}/\mathrm{days}$ & $P_\mathrm{orb}^\mathrm{min}/\mathrm{days}$ & $P_\mathrm{orb}^\mathrm{max}/\mathrm{days}$ & $t_\mathrm{pRGB}^\mathrm{mean}/\mathrm{kyr}$ & $t_\mathrm{pRGB}^\mathrm{min}/\mathrm{kyr}$ & $t_\mathrm{pRGB}^\mathrm{max}/\mathrm{kyr}$ & $t_\mathrm{pRGB}^\mathrm{mean}/\mathrm{kyr}$ & $t_\mathrm{pRGB}^\mathrm{min}/\mathrm{kyr}$ & $t_\mathrm{pRGB}^\mathrm{max}/\mathrm{kyr}$ \\
& & & & & $Z=0.02$ & $Z=0.02$ & $Z=0.02$ & $Z=0.00142$ & $Z=0.00142$ & $Z=0.00142$ \\
\hline
J045745.46-683724.1 & LMC & 41 & 40 & 93 & 2000 & 700 & 2000 & 2000 & 500 & 2000 \\
J054144.88-712351.7 & LMC & 44 & 44 & 76 & 2000 & 700 & 2000 & 0 & 0 & 1000 \\
J053811.63-680824.9 & LMC & 46 & 39 & 59 & 2000 & 1000 & 2000 & 800 & 0 & 1000 \\
J053030.78-701805.5 & LMC & 48 & 47 & 64 & 1000 & 600 & 1000 & 0 & 0 & 0 \\
J051412.13-702026.0 & LMC & 54 & 48 & 122 & 2000 & 500 & 2000 & 1000 & 300 & 1000 \\
J051751.41-713507.3 & LMC & 58 & 34 & 105 & 1000 & 600 & 3000 & 1000 & 500 & 3000 \\
J045736.84-705127.2 & LMC & 63 & 47 & 77 & 1000 & 1000 & 2000 & 1000 & 1000 & 2000 \\
J052107.90-714104.4 & LMC & 63 & 39 & 102 & 1000 & 700 & 3000 & 1000 & 600 & 3000 \\
J045058.16-671634.4 & LMC & 76 & 59 & 231 & 1000 & 200 & 1000 & 700 & 200 & 900 \\
(...) & (...) & (...) & (...) & (...) & (...) & (...) & (...) & (...) & (...) & (...) \\
\hline
\end{tabular}
\tablefoot{The table lists the following information for each object: its identification, host Magellanic Cloud, estimated orbital period $P_\mathrm{orb}$ determined from its luminosity (see Appendix \ref{single star estimation}), and the estimated post-mass-transfer age $t_\mathrm{pRGB}$ determined from its position in the HRD for both metallicities (see Sect. \ref{ages}). $P_\mathrm{orb}^\mathrm{min}$ and $P_\mathrm{orb}^\mathrm{max}$ correspond to the orbital periods estimated using the luminosity upper and lower limits, respectively. We do not estimate the orbital periods of OGLE-LMC-T2CEP-032 and OGLE-LMC-T2CEP-200, as these have already been photometrically determined. The whole table is available at CDS.
\tablefoottext{a}{Lower limit on the maximum orbital period, as the luminosity upper limit of the object is larger than the maximum luminosity of the binary models.}
\tablefoottext{b}{Lower limit on the orbital period, as the luminosity of the object is larger than the maximum luminosity of the binary models.}
\tablefoottext{c}{Upper limit on the minimum post-mass-transfer age, as the luminosity upper limit of the object is larger than the maximum luminosity of the binary models.}
\tablefoottext{d}{Upper limit on the mean post-mass-transfer age, as the luminosity of the object is larger than the maximum luminosity of the binary models.}
\tablefoottext{e}{Upper limit on the minimum post-mass-transfer age, as the luminosity lower limit of the object is larger than the maximum luminosity of the binary models.}
}
\end{sidewaystable*}

\begin{table*}
\caption{Estimated post-mass-transfer ages of the Galactic post-RGB binary subsample.}
\label{table:Galacticages}
\centering
\begin{tabular}{c c c c c c c}
\hline\hline
Name & $t_\mathrm{pRGB}^\mathrm{mean}/\mathrm{kyr}$ & $t_\mathrm{pRGB}^\mathrm{min}/\mathrm{kyr}$ & $t_\mathrm{pRGB}^\mathrm{max}/\mathrm{kyr}$ & $t_\mathrm{pRGB}^\mathrm{mean}/\mathrm{kyr}$ & $t_\mathrm{pRGB}^\mathrm{min}/\mathrm{kyr}$ & $t_\mathrm{pRGB}^\mathrm{max}/\mathrm{kyr}$ \\
& $Z=0.02$ & $Z=0.02$ & $Z=0.02$ & $Z=0.00142$ & $Z=0.00142$ & $Z=0.00142$ \\
\hline
EZ Gem & 4000 & 3000 & 10000 & 3000 & 1000 & 10000 \\
AU Peg & 2000 & 2000 & 3000 & 1000 & 700 & 2000 \\
EF Vul & 0 & 0 & 600 & 0 & 0 & 0 \\
IRAS 13258-8103 & 1000 & 100 & 9000 & 600 & 70 & 6000 \\
HD 213985; HM Aqr & 1000 & 600 & 1000 & 900 & 500 & 1000 \\
SZ Mon & 500 & 400 & 800 & 0 & 0 & 100 \\
BT Lac & 500 & 200 & 800 & 200 & 0 & 400 \\
IRAS 22251+5406 & 200 & 200 & 300 & 200 & 100 & 200 \\
IRAS 06165+3158 & 200 & 100 & 200 & 200 & 100 & 200 \\
HD 186438 & 200 & 100 & 200 & 100 & 100 & 200 \\
AY Lep & 90 & 0 & 100 & 0 & 0 & 0 \\
R Sge & 200 & 90 & 200 & 100 & 70 & 100 \\
CC Lyr & 200 & 100 & 200 & 100 & 80 & 200 \\
DY Ori & 200 & 100 & 200 & 100 & 80 & 200 \\
DF Cyg & 100 & 100 & 200 & 70 & 0 & 100 \\
IRAS 06452-3456 & 100 & 60\tablefootmark{a} & 500 & 100 & 60\tablefootmark{a} & 300 \\
UY Ara & 100 & 80 & 200 & 100 & 60 & 100 \\
SU Gem & 100 & 90 & 200 & 100 & 30 & 100 \\
ST Pup & 100 & 100 & 200 & 100 & 80 & 100 \\
HD 340949 & 100 & 90 & 300 & 70 & 0 & 200 \\
AR Pup & 100 & 60\tablefootmark{a} & 200 & 80 & 60\tablefootmark{a} & 100 \\
EZ Aql & 100 & 90 & 300 & 70 & 70 & 200 \\
GK Car & 100 & 80 & 200 & 90 & 60 & 100 \\
V1504 Sco & 100 & 80 & 300 & 90 & 70 & 200 \\
SX Cen & 100 & 90 & 100 & 90 & 30 & 100 \\
V Vul & 100 & 60\tablefootmark{a} & 200 & 70 & 50\tablefootmark{a} & 100 \\
IRAS 20094+3721 & 100 & 80 & 100 & 80 & 60\tablefootmark{a} & 100 \\
V1711 Sgr & 90 & 70 & 100 & 60 & 40 & 80 \\
TW Cam & 80 & 70 & 100 & 20 & 0 & 70 \\
CN Cen & 80 & 50\tablefootmark{a} & 100 & 60 & 40\tablefootmark{a} & 90 \\
GZ Nor & 70 & 50\tablefootmark{a} & 100 & 40 & 0 & 80 \\
GP Cha & 80 & 60\tablefootmark{a} & 100 & 60 & 50\tablefootmark{a} & 80 \\
V1333 Sco & 80 & 60\tablefootmark{a} & 90 & 60\tablefootmark{b} & 30\tablefootmark{a} & 60 \\
RV Tau & 70 & 50\tablefootmark{a} & 80 & 20\tablefootmark{b} & 0 & 50 \\
RS Sge & 80 & 60\tablefootmark{a} & 100 & 60\tablefootmark{b} & 60\tablefootmark{a} & 90 \\
HD 114855; V956 Cen & 60 & 60\tablefootmark{a} & 90 & 60\tablefootmark{b} & 60\tablefootmark{a} & 70 \\
HD 137569 & 60\tablefootmark{b} & 60\tablefootmark{a} & 90 & 70\tablefootmark{b} & 70\tablefootmark{a} & 80 \\
\hline
\end{tabular}
\tablefoot{The table lists the following information for each object: its name and estimated mean, minimum, and maximum post-mass-transfer age $t_\mathrm{pRGB}$ determined from its position in the HRD for both metallicities (see Sect. \ref{relation}).
\tablefoottext{a}{Upper limit on the minimum post-mass-transfer age, as the luminosity upper limit of the object is larger than the maximum luminosity of the binary models.}
\tablefoottext{b}{Upper limit on the mean post-mass-transfer age, as the luminosity of the object is larger than the maximum luminosity of the binary models.}
}
\end{table*}

\end{appendix}

\end{document}